\documentclass[prd,twocolumn,nofootinbib]{revtex4-1}

\usepackage{multirow}
\usepackage{amsfonts}
\usepackage{amsmath}
\usepackage{amssymb}
\usepackage{bm}
\usepackage{dcolumn}
\usepackage{graphicx}
\usepackage{graphics}
\usepackage[latin1]{inputenc}
\usepackage{latexsym}
\usepackage{rotating}
\usepackage[colorlinks=true]{hyperref}
\usepackage[all]{hypcap} 
\usepackage{xspace} 
\usepackage[usenames]{color}
\usepackage{mathrsfs}
\usepackage{subfigure}

\usepackage{booktabs}
\usepackage{tabu}
\usepackage{ulem}
\newcommand\be{\begin{equation}}
\newcommand\ba{\begin{eqnarray}}
\newcommand\ee{\end{equation}}
\newcommand\ea{\end{eqnarray}}
\newcommand\bw{\begin{widetext}}
\newcommand\ew{\end{widetext}}

\newcommand{\tol}{{\mbox{\tiny Tol}}}

\begin{document}
\title{Improved Analytic Modeling of Neutron Star Interiors}    

\author{Nan Jiang}
\affiliation{Department of Physics, University of Virginia, Charlottesville, Virginia 22904, USA.}

\author{Kent Yagi}
\affiliation{Department of Physics, University of Virginia, Charlottesville, Virginia 22904, USA.}

\begin{abstract} 

Studies of neutron stars are extremely timely given the recent detection of gravitational waves from a binary neutron star merger GW170817, and an International Space Station payload NICER currently in operation that aims to determine radii of neutron stars to a precision better than 5\%. In many cases, neutron star solutions are constructed numerically due to the complexity of the field equations with realistic equations of state. However, in order to relate observables like the neutron star mass and radius to interior quantities like central density and pressure, it would be useful to provide an accurate, analytic modeling of a neutron star interior. One such solution for static and isolated neutron stars is the Tolman VII solution characterized only by two parameters (e.g. mass and radius), though its agreement with numerical solutions is not perfect. We here introduce an improved analytic model based on the Tolman VII solution by introducing an additional parameter to make the analytic density profile agree better with the numerically obtained one. This additional parameter can be fitted in terms of the stellar mass, radius and central density in an equation-of-state-insensitive way. In most cases, we find that the new model more accurately describes realistic profiles than the original Tolman VII solution by a factor of 2--5. Our results are first-step calculations towards constructing analytic interior solutions for more realistic neutron stars under rotation or tidal deformation.

\end{abstract}

\date{\today}
\maketitle
%

\section{Introduction}
\label{sec:intro}

Studies of neutron stars (NSs) can bring valuable information about fundamental physics, including nuclear physics. NSs consist of matter with densities that exceed nuclear saturation density. Thus they offer  natural laboratories to probe nuclear matter equation of state (EoS) (relation between pressure and energy density) that is difficult to access with ground-based nuclear experiments~\cite{0004-637X-550-1-426,lattimer-prakash-review,Ozel:2016oaf}. One way to extract internal structure information is to measure the NS mass and radius independently~\cite{ozel-baym-guver,steiner-lattimer-brown,Ozel:2015fia}, though current measurements may contain large systematic errors. An X-ray astrophysics payload NICER currently in operation at the International Space Station is expected to measure the stellar radius to $\sim 5\%$ accuracy~\cite{Ozel:2015ykl} with less systematics~\cite{Lo:2013ava,Lo:2018hes}. Another way to probe internal structure is to measure tidal deformabilities of neutron stars via gravitational waves. The recent event GW170817 favors softer EoSs~\cite{PhysRevLett.119.161101,Abbott:2018wiz,Abbott:2018exr} that tend to produce NSs with smaller radii and maximum masses. GW170817 can also be used to infer nuclear parameters around saturation density~\cite{Malik:2018zcf,Carson:2018xri}. Neutron stars are also useful to probe General Relativity, as evidenced by binary pulsar~\cite{GRtest_pulsar,stairs} and gravitational wave~\cite{Monitor:2017mdv,Abbott:2018lct} observations.

In order to connect NS observables (masses, radii, tidal deformabilities etc.) to internal structure, one needs to construct NS solutions by solving the Einstein equations with a given EoS. Most of such solutions are constructed numerically due to the complex nature of the field equations. Having said this, analytic solutions to the Einstein equations exist that can mimic realistic NS solutions. One simplest example is a solution with constant density (Schwarzschild interior solution)~\cite{Schutz:1985jx}. Analytic solutions for modeling more realistic stars include Buchdahl~\cite{1967ApJ...147..310B,0004-637X-550-1-426,Schutz:1985jx} and Tolman VII~\cite{PhysRev.55.364,0004-637X-550-1-426,Raghoonundun:2015wga} solutions. The latter is stable for a large range of compactness~\cite{TolmanVIIstability} and its geometric structures were studied in~\cite{Neary:2001ai,Raghoonundun:2016cun}.

Analytic NS solutions are useful to have a better understanding of NS physics. 
NS quasinormal modes and associated universal relations have been investigated in detail with the Tolman VII solution~\cite{PhysRevLett.95.151101,0004-637X-631-1-495,Tsui:2006tr}. An analytic constant density solution with anisotropic pressure~\cite{Bowers:1974tgi} was used to study how universal relations between moment of inertia ($I$), tidal Love number and quadrupole moment ($Q$), and hence I-Love-Q relations, approach the black hole limit~\cite{Yagi:2016ejg}. These analytic solutions for NSs can also be useful to examine non-GR theories. For example, constant density and Tolman VII solutions were used to investigate how stellar scalar charges vanish in string-inspired theories of gravity~\cite{Yagi:2015oca}.

In this paper, we begin by comparing the Tolman VII solution with numerical solutions. The density profile among these solutions were investigated in~\cite{0004-637X-550-1-426}. Here, we also study the profiles for the interior mass, gravitational potential and pressure. For a 1.4$M_\odot$ NS with the AP4 EoS (a soft EoS consistent with the LIGO-Virgo tidal measurement~\cite{PhysRevLett.119.161101,Abbott:2018wiz,Abbott:2018exr}), the density and mass profiles for the Tolman solution match with the numerical ones with a typical error of $\sim 10$\%. 

The main goal of this paper is to find an analytic model of the NS interior that can more accurately describe the realistic solution obtained numerically than the original Tolman VII solution. The latter models the density to be a quadratic function of the radial coordinate $r$. We here introduce an additional parameter $\alpha$ to allow the density to be a quartic function of $r$. We find an approximate universal relation among this additional parameter and the stellar mass $M$, radius $R$ and central density $\rho_c$ that is insensitive to the underlying EoS. The final expression is a three-parameter solution in terms of $M$, $R$ and $\rho_c$. The price one has to pay by introducing the additional parameter is that the density profile is slightly more complicated than the original model and we could not find an \emph{exact} analytic solution to the Einstein equations. 

Having said this, we managed to find an approximate, three-parameter solution that can more accurately model realistic profiles than the original Tolman VII solution in most cases. For example, the density and mass profiles of a 1.4$M_\odot$ NS with the AP4 EoS now agree with the numerical ones within an error of $\sim 1\%$. Regarding other masses and EoSs,  the new model can more accurately model numerical results compared to the original Tolman solution by a factor of 2--5.  The new model outperforms the original one especially for softer EoSs with a relatively large mass ($> 1.5 M_\odot$). The accuracy of the new model can be improved further if we use 
a fit for $\alpha$ that is specific to each EoS, though the improvement from the case with the universal-$\alpha$ fit is not so significant. 

The remaining of the paper is organized as follows. In Sec.~\ref{sec:Original Tolman}, we review the original Tolman VII solution  while in Sec.~\ref{sec:Improved Tolman}, we present our new model.
In Sec.~\ref{sec:Comparision}, we compare the two models and show that the new model has a better agreement with numerical solutions than the original model in most cases, especially for softer EoSs. We conclude in Sec.~\ref{sec:Conclusion} and give possible directions for future work.
For busy readers, we summarize the original and improved Tolman VII solutions in Table~\ref{table:summary}.
We use the geometric units of $c=1$ and $G=1$ throughout this paper unless otherwise stated.

\section{Original Tolman VII Solution}
\label{sec:Original Tolman}

We begin by reviewing the original Tolman VII solution~\cite{PhysRev.55.364} that can mimic static and spherically symmetric NSs~\cite{0004-637X-550-1-426}. We use the metric ansatz given by
\ba
\label{eq:metric}
ds^2 = -e^\nu dt^2+ e^\lambda dr^2+r^2(d\theta^2+\sin^2\theta d\phi^2).
\ea
Here, 
$\nu$ and $\lambda$ are functions of $r$ only.
We assume matter inside a NS can be modeled by a perfect fluid whose stress-energy tensor is given by
\ba
\label{eq:matter}
T_{\mu\nu}=(\rho+p) u_\mu u_\nu+ p g_{\mu\nu},
\ea
where $u^\mu$ is the four-velocity of the fluid while $\rho$ and $p$ represent the matter energy density and pressure respectively.

Substituting Eqs.~\eqref{eq:metric} and~\eqref{eq:matter} into the Einstein equations, one finds independent equations as~\cite{PhysRev.55.364} 
\be
\label{eq:diff-eq-nu}
\frac{d}{dr} \left(\frac{e^{-\lambda}-1}{r^2}\right)+\frac{d}{dr}\left( \frac{e^{-\lambda} \nu'}{2 r}\right)+e^{-\lambda-\nu} \frac{d}{dr}\left (\frac{e^\nu \nu'}{2 r}\right) = 0,
\ee
\ba
e^{-\lambda} \left(\frac{\nu'}{r}+\frac{1}{r^2}\right)-\frac{1}{r^2} =8 \pi p,
\label{pressure formula}
\ea
\ba
\label{eq:m}
\frac{dm}{dr} = 4\pi r^2 \rho, 
\ea
where a prime denotes a derivative with respect to $r$ and
\be
e^{-\lambda} \equiv 1 - \frac{2m}{r}\,. 
\ee
 To close the system of equations, one normally chooses an EoS that relates $p$ as a function of $\rho$.

Instead of choosing an EoS, Tolman specified $e^{-\lambda}$ to be a quartic function of $r$. This leads to the energy density profile of
\ba
\label{eq:rho-Tol}
\rho_\tol(r) = \rho_{c} (1-\xi^2),
\ea
where $\xi=r/R$ with $R$ representing the stellar radius and $\rho_c$ being the central energy density. The subscript ``Tol'' refers to the quantity in the original Tolman solution. Substituting this into Eq.~\eqref{eq:m} and integrating over $r$ with the boundary condition $m(0)=0$, one finds
\ba
\label{eq:m-Tol}
m_\tol(r)  =  4 \pi \rho_{c} \left(\frac{r^3}{3} - \frac{r^5}{5 R^2}\right).
\ea
$\rho_c$ can be expressed in terms of the stellar mass $M \equiv m(R)$ as
\be
\rho_c = \frac{15 M}{8 \pi  R^3}.
\ee 
Substituting this back into Eqs.~\eqref{eq:rho-Tol} and~\eqref{eq:m-Tol}, one finds
\ba
\label{eq:original Tol}
\rho_\tol(r) & =& \frac{15 M}{8 \pi R^3}  (1- \xi^2), \\
\label{original m}
m_\tol(r) &=& M \left(\frac{5}{2} \xi^3 - \frac{3}{2}\xi^5\right).
\ea
$e^{-\lambda}$ is given by a quartic polynomial in terms of $\xi$ as
\ba
e^{-\lambda_\tol(r)} &=& 1- \mathcal{C} \xi^2 \left(5 - 3 \xi^2\right) \\
&=& 1- \frac{8 \pi}{15} R^2\rho_c \xi^2 \left(5 - 3 \xi^2\right),
\label{exp(-l) expression}
\ea
where 
\be
\mathcal{C} \equiv \frac{M}{R} = \frac{8 \pi}{15} R^2\rho_c
\ee
is the stellar compactness.

With these expressions at hand, Tolman~\cite{PhysRev.55.364} analytically solved for $\nu$ and $p$. First, Eq.~\eqref{eq:diff-eq-nu} can be integrated to yield
\ba
e^{\nu_\tol(r)} = C_1^\tol \cos^2\phi_\tol,
\label{original nu}
\ea
with 
\ba
\phi_\tol &=& C_2^\tol - \frac{1}{2} \log\left(\xi^2 - \frac{5}{6} +\sqrt{\frac{e^{-\lambda_\tol}}{3 \mathcal{C}}}\right) \nonumber \\
\label{eq:phi_orig}
 &=& C_2^\tol - \frac{1}{2} \log\left(\xi^2 - \frac{5}{6} +\sqrt{\frac{5 e^{-\lambda_\tol}}{8 \pi R^2 \rho_c}}\right).
\ea
The integration constants $C_1^\tol$ and $C_2^\tol$ are determined from the boundary conditions
\ba
e^{\nu_\tol(R)} = 1-\frac{2 M}{R}, \qquad p_\tol(R)=0\,,
\label{first boundary condition} 
\ea
with $p_\tol$ given by Eq.~\eqref{pressure formula}.
One finds
\ba
C_1^\tol &=& 1 - \frac{5 \mathcal{C}}{3}, \\
C_2^\tol &=& \arctan \sqrt{\frac{\mathcal{C}}{3 (1-2 \mathcal{C})}} + \frac{1}{2}\log\left(\frac{1}{6} + \sqrt{\frac{1 - 2\mathcal{C}}{3 \mathcal{C}}}\right), \nonumber \\
\ea
and
\ba
p_\tol = \frac{1}{4 \pi R^2} \left[ \sqrt{3 \mathcal{C} e^{- \lambda}} \tan\phi_\tol- \frac{\mathcal{C}}{2} (5 - 3 \xi^2)\right].
\label{original p}
\ea
The above solution is the so-called Tolman VII solution.

\begin{figure}[htb]
\includegraphics[width=8.5cm]{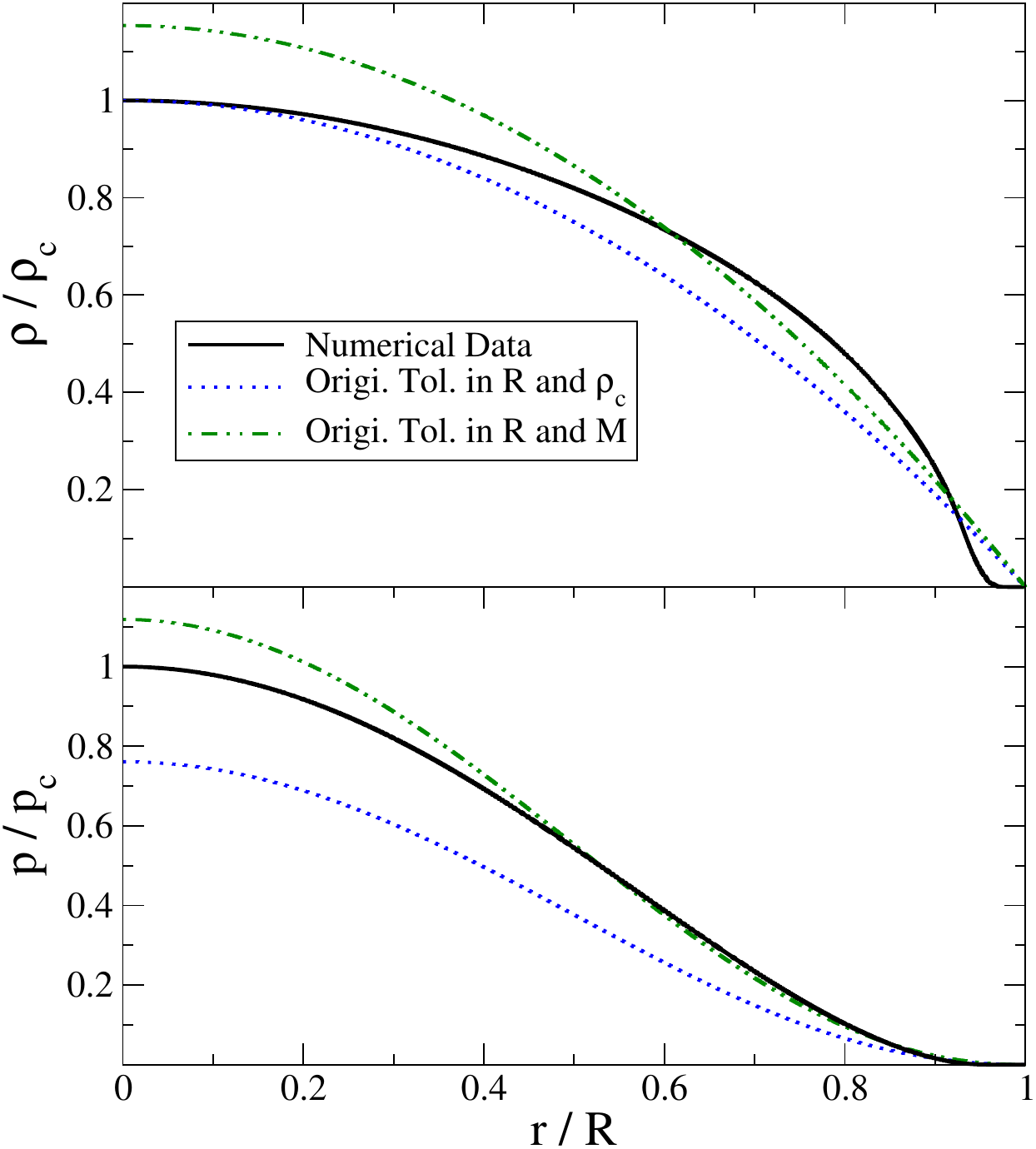}
\caption{Energy density (top) and pressure (bottom) profiles for the original Tolman solution with two different parameterizations. We choose $R=11.4$km and either $\rho_c=9.9\times10^{14}$g/cm$^3$ or $M=1.4M_\odot$. We also present the numerical solution with the AP4 EoS and $\rho_c=9.9\times10^{14}$g/cm$^3$ that corresponds to $M=1.4M_\odot$ and $R=11.4$km. }
\label{fig:p_rho_Tol}
\end{figure}

Figure~\ref{fig:p_rho_Tol} presents the normalized energy density and pressure profiles of a 1.4$M_\odot$ NS with the AP4 EoS for two different parameterizations of the original Tolman solution. For reference we also show realistic profiles obtained numerically. Regarding the energy density profile, observe that the $(R,\rho_c)$ parameterization more accurately models the realistic profile near the stellar center. This is because the central density is a free parameter that we can choose to be the value that matches the one with the numerical calculation.  On the other hand, the $(R,M)$ parameterization works better in the intermediate regime of the star. We found a similar feature for the $m$ profile as it is obtained simply by integrating $\rho$ over a volume as in Eq.~\eqref{eq:m}. 

Regarding the pressure profile, the $(R,M)$ parametrization works better throughout, and we found a similar feature for the $\nu$ profile. This is because $p$ is obtained from $\nu$ (see Eq.~\eqref{pressure formula}), which is determined from the boundary condition at the stellar surface in terms of $R$ and $M$ (Eq.~\eqref{first boundary condition}). Thus, the $(R,M)$ parameterization allows one to match $\nu$ at the surface perfectly with the numerical value. This suggests that perhaps the $(R,M)$ parameterization has more advantage than the $(R,\rho_c)$ one, except near the center of the $\rho$ and $m$ profiles.  

\section{Improved Tolman VII Modeling}
\label{sec:Improved Tolman}

We here propose an improved model which has three free parameters $(M, R,\rho_c)$. We begin by introducing an additional term to Eq.~\eqref{eq:rho-Tol}:
\ba
\label{eq:rho_imp}
\rho_\mathrm{imp}(r) = \rho_{c} \left[1- \alpha \xi^2 + (\alpha - 1) \xi^4\right],\label{mod rho}
\ea
with a constant $\alpha$. The coefficients are chosen such that $\rho_\mathrm{imp}(R) =0$. The original Tolman solution is recovered in the limit $\alpha \to 1$. $m$ and $\lambda$ now become
\ba
m_\mathrm{imp} &=& 4 \pi \rho_c R^3 \xi^3 \left(\frac{1}{3} - \frac{\alpha}{5} \xi^2 +\frac{\alpha - 1}{7} \xi^4\right), \\
e^{-\lambda_\mathrm{imp}} &=& 1- 8 \pi R^2 \xi^2 \rho_c \left(\frac{1}{3} -\frac{\alpha}{5} \xi^2+ \frac{\alpha-1}{7} \xi^4 \right).
\label{mod m}
\ea

\subsection{Choice of $\alpha$ }

Before deriving the improved expression for $\nu$ and $p$, let us see how we can express $\alpha$ in terms of $M$, $R$ and $\rho_c$.
One way to determine this is to use $M = m_\mathrm{imp}(R)$, which yields
\ba
\label{eq:alpha_analytic}
\alpha = \frac{5 (-21  M + 16 \pi R^3 \rho_c )}{24 \pi R^3 \rho_c}.
\ea
However, we find that a more accurate modeling is obtained by fitting Eq.~\eqref{mod rho} to the true density profile obtained numerically for various EoSs and $\rho_c$. We adopt eleven realistic EoSs with different stiffness as summarized in Table~\ref{tab:EoS class}. These EoSs all support a $2M_\odot$ NS~\cite{2.01NS}.
We consider fits for $\alpha$ in terms of $M$, $R$ and $\rho_c$ given by
\ba
\alpha = a_0 + a_1 \left( \frac{\mathcal{C}^{n}}{\rho_c R^2} \right) + a_2 \left(\frac{\mathcal{C}^{n}}{\rho_c R^2} \right)^2,
\label{alpha expression}
\ea
where $\mathcal{C}=M/R$ and the fitted coefficients $a_0$, $a_1$, $a_2$ and $n$ for each EoS are summarized in Table~\ref{Tab: alpha}.

\begin{figure}[htb]
\includegraphics[width=8.5cm]{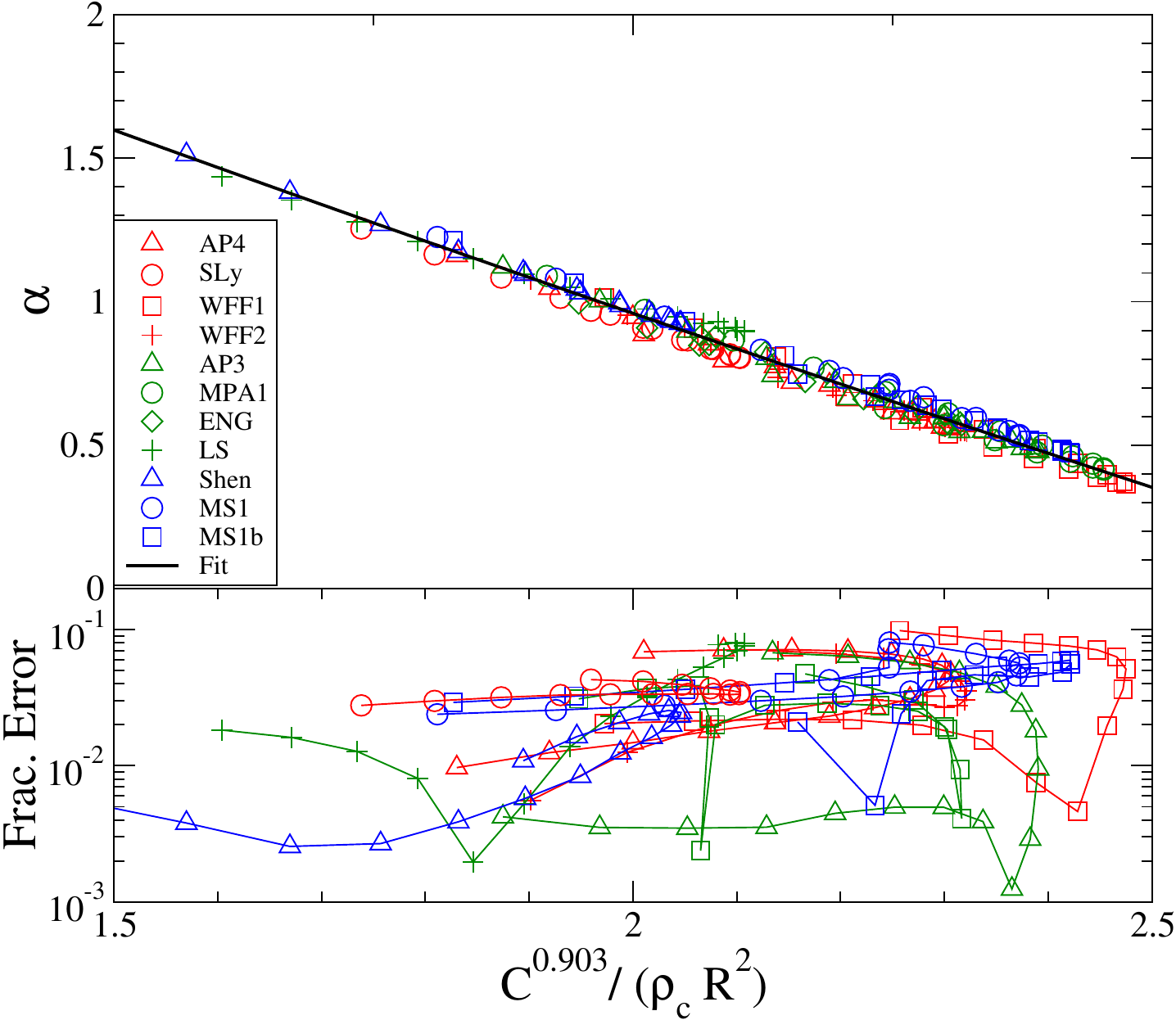}
\caption{ (Top)  $\alpha$ (characterizing the density profile in the improved Tolman model in Eq.~\eqref{eq:rho_imp}) as a function of $C^n/\rho_c R^2$ with $n=0.903$ for 11 realistic EoSs. Different colors correspond to different classes of EoSs in Table~\ref{tab:EoS class} (soft in red, intermediate in green and stiff in blue). We also present the fit in a black solid curve given by Eq.~\eqref{alpha expression} with the coefficients given in the last row of Table~\ref{Tab: alpha}.
 (Bottom) Relative fractional errors between numerical results and the fit. Notice that the relations are nearly EoS independent, with an EoS-variation of 10\% at most. 
}
\label{fig:alpha}
\end{figure}

Such EoS-specific fits for $\alpha$ are useful only if one wishes to model the NS interior solution accurately for the EoSs presented in Table~\ref{Tab: alpha}, and perhaps it would be more useful if we have a single, universal fit for $\alpha$ that is valid for any EoSs. The top panel of Fig.~\ref{fig:alpha} shows $\alpha$ against $\mathcal{C}^n/\rho_c R^2$ with $n=0.903$ for various EoSs. Indeed, the relation seems to be universal in the sense that it is insensitive to the choice of EoS. Based on this finding, we created a single fit, again using Eq.~\eqref{alpha expression}, that is valid for all 11 EoSs considered here. The fitting coefficients are summarized in Table~\ref{Tab: alpha}. 

The bottom panel of Fig.~\ref{fig:alpha} presents the fractional difference between each data point and the universal fit. Observe that the fit is valid to 10\% accuracy for any EoSs. 
Note that as one increases the central density, each sequence reaches a maximum value for $\mathcal{C}^n/\rho_c R^2$ and starts to turn around. This leads to the fact that the fractional difference from the fit being larger for larger $\mathcal{C}^n/\rho_c R^2$. In such a region, there can be two different values for $\alpha$ for a fixed $\mathcal{C}^n/\rho_c R^2$ (again due to the turn over), and thus it becomes more difficult to fit the relation.

\renewcommand{\arraystretch}{1.3}
\begingroup
\begin{table}[h]
\begin{centering}
\begin{tabular}{l |  l}
\cline{1-2}
EoS class    &Members \\
\hline
soft    & AP4~\cite{AP3}, SLy~\cite{SLy}, WFF1~\cite{WFF1}, WFF2~\cite{WFF1}\\
intermediate      & ENG~\cite{ENG}, MPA1~\cite{MPA1}, AP3~\cite{AP3}, LS~\cite{LS}\\
stiff      &Shen~\cite{SHEN1998435}, MS1~\cite{MS1}, MS1b~\cite{MS1}\\
\hline
\end{tabular}
\end{centering}
\caption{Eleven realistic EoSs considered in this paper. They are categorized into three different stiffness classes ~\cite{EoSclass}.}
\label{tab:EoS class}
\end{table}
\endgroup

\renewcommand{\arraystretch}{1.3}
\begingroup
\begin{table}[t]
\begin{centering}
\begin{tabular}{l | c c c c | c}
\cline{1-6}
EoS    & $a_0$ & $a_1$ & $a_2$ & $ n $ & R-squared \\
\hline
AP4     & 3.90061   &-1.67716 & 0.112974 &0.884655  &1.000000\\
SLy      & 4.08125     &-1.94944  & 0.190047  &0.898685 &1.000000\\
WFF1  & 3.49902  &-1.24206  &0.01264 &0.871133 &0.999996\\
WFF2  &5.00228  &-2.70395 &0.347978 &0.88916 &0.999998\\
AP3 &3.99892 &-1.75538 &0.133497 &0.881961 &1.000000\\
MPA1  &3.84739 &-1.58061 &0.0919565 &0.879148 &0.999999\\
ENG &0.438372 &1.28922 &-0.506597 &0.874422 &0.999733\\
LS        & 4.18945   &-2.20875  &0.288819 &0.920735   &1.000000\\
Shen    & 4.05847    &-1.92481  &0.187936 &0.906579   &0.999998\\
MS1   &3.74656 &-1.51608 &0.0612786 &0.911464 &0.999909\\
MS1b  &3.95158 &-1.69133 &0.114453 &0.891669 &0.999914\\
\hline
universal     & 3.70625   &-1.50266 & 0.0643875 &0.903  &0.998772\\
\hline
\end{tabular}
\end{centering}
\caption{
Fitted coefficients of $\alpha$ in Eq.~\eqref{alpha expression} for each realistic EoS. We also present a universal fit for $\alpha$ that is valid for all the realistic EoSs considered here within an error of $10\%$. The last column shows 
the R-squared value that gives a statistical measure of how good the fit is. It is the coefficient of determination defined by $\mathrm{R^2} \equiv 1-\frac{\sum_{i} (\alpha_i-\bar{\alpha})^2}{\sum_{i} (\alpha_i- f_i)^2}$, in which $\alpha_i $ represents the numerical data, $f_i$ is the predicted value from the model and $\bar{\alpha}$ is the mean of the numerical data.
}
\label{Tab: alpha}
\end{table}
\endgroup

\renewcommand{\arraystretch}{1.8}
\begingroup
\begin{table}[th]
\begin{centering}
\begin{tabular}{ c | c }
\hline
&  $\rho_\tol = \frac{15\mathcal{C}}{8 \pi R^2}  (1- \xi^2)$ \\
 &  $m_\tol = R \, \mathcal{C} \left(\frac{5}{2} \xi^3 - \frac{3}{2}\xi^5\right)$ \\
 & $e^{\nu_\tol} = C_1^\tol \cos^2\phi_\tol$ \\
 Original&
$p_\tol = \frac{1}{4 \pi R^2} \left[ \sqrt{3 \mathcal{C}e^{-\lambda_\tol}} \tan\phi_\tol- \frac{\mathcal{C}}{2} (5 - 3 \xi^2)\right]$ \\ 
 Tolman& $\phi_\tol = C_2^\tol - \frac{1}{2} \log\left(\xi^2 - \frac{5}{6} +\sqrt{\frac{e^{-\lambda_\tol}}{3 \mathcal{C}}}\right)$ \\
 & $C_1^\tol=1 - \frac{5 \mathcal{C}}{3}$ \\
 & $C_2^\tol=\arctan \sqrt{\frac{\mathcal{C}}{3 (1-2 \mathcal{C})}} + \frac{1}{2}\log\left(\frac{1}{6} + \sqrt{\frac{1 - 2\mathcal{C}}{3 \mathcal{C}}}\right)$ \\
 \hline
& $\rho_\mathrm{imp} = \rho_{c} \left[1- \alpha \xi^2 + (\alpha - 1) \xi^4\right]$ \\
 & $m_\mathrm{imp} =4 \pi \rho_c R^3 \left(\frac{\xi^3}{3} - \frac{\alpha \xi^5}{5} +\frac{\alpha - 1}{7} \xi^7\right)$ \\
 & $e^{\nu_\mathrm{imp}}=C_1^\mathrm{imp} \cos^2\phi_\mathrm{imp}$ \\
 & $p_\mathrm{imp}=  \sqrt{\frac{e^{-\lambda_\tol} \text{$\rho $}_c}{10
   \pi }
   } \frac{\tan \phi_\mathrm{imp}}{R}
+\frac{1}{15} \left(3 \text{$\xi
   $}^2-5\right) \text{$\rho $}_c$ \\
 Improved  & $+\frac{6
   (1-\alpha ) \text{$\rho $}_c}{16 \pi  (10-3
   \alpha) \text{$\rho $}_c
   R^2-105}$ \\
Tolman & $\phi_\mathrm{imp}=C_2^\mathrm{imp} - \frac{1}{2} \log\left(\xi^2 - \frac{5}{6} +\sqrt{\frac{5 e^{-\lambda_\tol}}{8 \pi R^2 \rho_c}}\right)$ \\
 & $C_1^\mathrm{imp}=(1-2\mathcal{C})  \left\{ 1+\frac{8 \pi R^2 \rho_c (10 -3\alpha)^2 (15 - 16 \pi R^2 \rho_c)}{3 [105+16\pi R^2  \rho_c (3\alpha-10)]^2}\right\}$ \\
 & $C_2^\mathrm{imp}=\arctan \left[-\frac{2 (10-3 \alpha )
  R \sqrt{6 \pi  \text{$\rho $}_c  \left(15-16 \pi 
   \text{$\rho $}_c R^2\right)}}{48 \pi
    (10-3 \alpha) \text{$\rho $}_c
   R^2-315}\right]$ 
 \\
 & $ + \frac{1}{2}\log\left(\frac{1}{6} + \sqrt{\frac{5}{8 \pi  \text{$\rho $}_c R^2}-\frac{2}{3}}\right)$ \\
 & $\alpha = a_0 + a_1 \left( \frac{\mathcal{C}^{n}}{\rho_c R^2} \right) + a_2 \left(\frac{\mathcal{C}^{n}}{\rho_c R^2} \right)^2$ \\
 \hline
\end{tabular}
\end{centering}
\caption{Summary of the original Tolman solution (top) and the improved model (bottom), with $\xi = r/R$. We present the energy density $\rho$, the interior mass $m$, the $(t,t)$ component of the metric $e^\nu (=-g_{tt})$ and the pressure $p$. The $(r,r)$ component of the metric is related to $m$ as $g_{rr} = e^{\lambda} = (1-2m/r)^{-1}$. Fitting coefficients $a_0$, $a_1$, $a_2$ and $n$ in $\alpha$ are summarized in Table~\ref{Tab: alpha}. The $(R,\rho_c)$ parameterization of the original Tolman solution is obtained by setting the stellar compactness as $\mathcal{C} = (8\pi/15) R^2 \rho_c$, while the $(R,M)$ parameterization of the original Tolman solution and the improved Tolman model uses $\mathcal{C} = M/R$. We stress that $\lambda$ entering in $p_\mathrm{imp}$ and $\phi_\mathrm{imp}$ is $\lambda_\tol$ and not $\lambda_\mathrm{imp}$. 
}
\label{table:summary}
\end{table}
\endgroup

\subsection{Improved Analytic Expressions for $\nu$ and $p$ }

Next, we look for the expressions for $\nu$ and $p$. The price we have to pay for adding the additional term in Eq.~\eqref{mod rho} is that we are no longer able to solve Eq.~\eqref{eq:diff-eq-nu} analytically. Thus, we find an approximate solution instead. 

Let us first derive the improved expression for $\nu$. We begin by approximating $\lambda$ in Eq.~\eqref{eq:diff-eq-nu} with the original Tolman VII expression $\lambda_\tol$ and not the improved version $\lambda_\mathrm{imp}$. The solution for $\nu$ to this equation then has the same form as Eqs.~\eqref{original nu} and~\eqref{eq:phi_orig}:
\be
e^{\nu_\mathrm{imp}(r)} = C_1^\mathrm{imp} \cos^2\phi_\mathrm{imp},
\ee
with
\ba
\phi_\mathrm{imp} &=& C_2^\mathrm{imp} - \frac{1}{2} \log\left(\xi^2 - \frac{5}{6} +\sqrt{\frac{5 e^{-\lambda_\tol}}{8 \pi R^2 \rho_c}}\right).
\ea
Though the integration constants $C_1^\mathrm{imp}$ and $C_2^\mathrm{imp}$ are different from the original ones $C_1^\tol$ and $C_2^\tol$ as we improve the boundary conditions: 
\ba
\label{eq:2nd-BC}
e^{\nu_\mathrm{imp}(R)} = 1-\frac{2 M}{R}, \qquad \bar p_\mathrm{imp}(R)=0\,.
\ea
These yield
\ba
C_1^\mathrm{imp}  &=& (1-2\mathcal{C})  \left\{ 1+\frac{8 \pi R^2 \rho_c (10 -3\alpha)^2 (15 - 16 \pi R^2 \rho_c)}{3 [105+16\pi R^2  \rho_c (3\alpha-10)]^2}\right\} \nonumber \\
   \\
C_2^\mathrm{imp} &=& \arctan \left[-\frac{2 (10-3 \alpha )
  R \sqrt{6 \pi  \text{$\rho $}_c  \left(15-16 \pi 
   \text{$\rho $}_c R^2\right)}}{48 \pi
    (10-3 \alpha) \text{$\rho $}_c
   R^2-315}\right], \nonumber \\
&&  + \frac{1}{2}\log\left(\frac{1}{6} + \sqrt{\frac{5}{8 \pi  \text{$\rho $}_c
   R^2}-\frac{2}{3}}\right).
\ea

Next, we derive the improved expression for $p$.
Using Eq.~\eqref{pressure formula}, the pressure for the improved model is given by
\be
\label{eq:pbar}
\bar p_\mathrm{imp} = \frac{1}{8\pi} \left[ e^{-\lambda_\mathrm{imp}} \left(\frac{\nu_\mathrm{imp}'}{r}+\frac{1}{r^2}\right)-\frac{1}{r^2}  \right].
\ee
However, we found that Eq.~\eqref{eq:pbar} gives the central pressure that is $\sim 20\%$ off from numerical results. Moreover, the pressure becomes negative near the surface, which is unphysical. These points can be remedied by changing $\lambda_\mathrm{imp}$ to $\lambda_\tol$ in Eq.~\eqref{eq:pbar} and shift the overall profile by a constant such that the pressure reduces to 0 at the surface:
\ba
p_\mathrm{imp} &= & \frac{1}{8\pi} \left[ e^{-\lambda_\tol(r)} \left(\frac{\nu_\mathrm{imp}'(r)}{r}+\frac{1}{r^2}\right)-\frac{1}{r^2}  \right] \nonumber \\
&& -  \frac{1}{8\pi} \left[ e^{-\lambda_\tol(R)} \left(\frac{\nu_\mathrm{imp}'(R)}{R}+\frac{1}{R^2}\right)-\frac{1}{R^2}  \right] \nonumber \\
&=& \sqrt{\frac{e^{-\lambda_\tol} \text{$\rho $}_c}{10
   \pi }
   } \frac{\tan \phi_\mathrm{imp}}{R}
+\frac{1}{15} \left(3 \text{$\xi
   $}^2-5\right) \text{$\rho $}_c \nonumber \\
   &&+\frac{6
   (1-\alpha ) \text{$\rho $}_c}{16 \pi  (10-3
   \alpha) \text{$\rho $}_c
   R^2-105}.
   \ea
The original Tolman solution and the improved model is summarized in Table~\ref{table:summary}.

We note that the set $(\rho_\mathrm{imp}, m_\mathrm{imp}, \nu_\mathrm{imp}, p_\mathrm{imp})$ is only an approximate solution to the Einstein equations. Having said this, $(\rho_\tol, m_\tol, \nu_\mathrm{imp}, p_\mathrm{imp})$ forms an exact solution to the Einstein equations, just like $(\rho_\tol, m_\tol, \nu_\tol, p_\tol)$. The difference between these two sets of exact solutions originates simply from different boundary conditions. The former uses Eq.~\eqref{eq:2nd-BC} while the latter adopts Eq.~\eqref{first boundary condition}.

\begin{figure*}[htb]
\includegraphics[width=8.5cm]{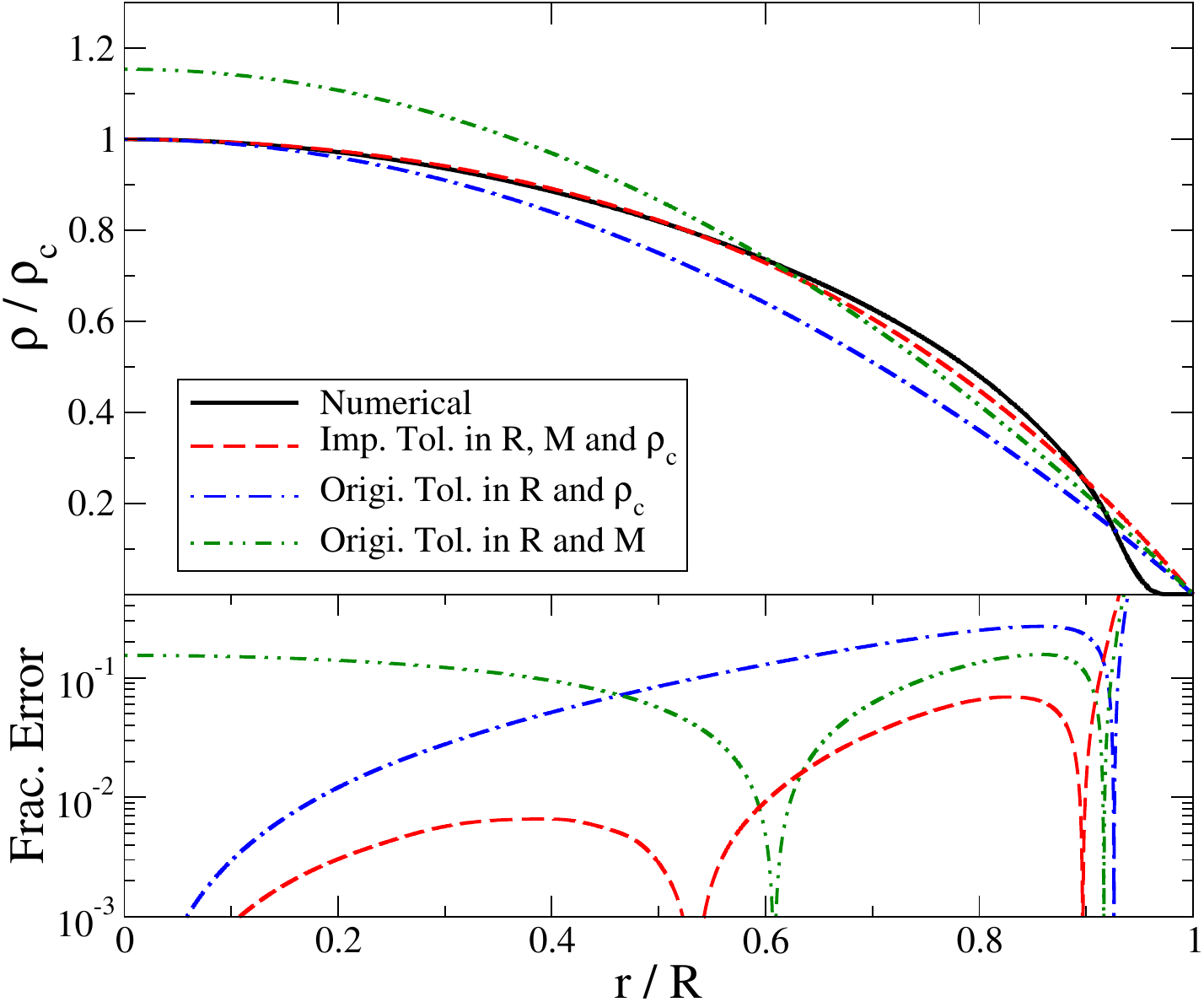}
\includegraphics[width=8.5cm]{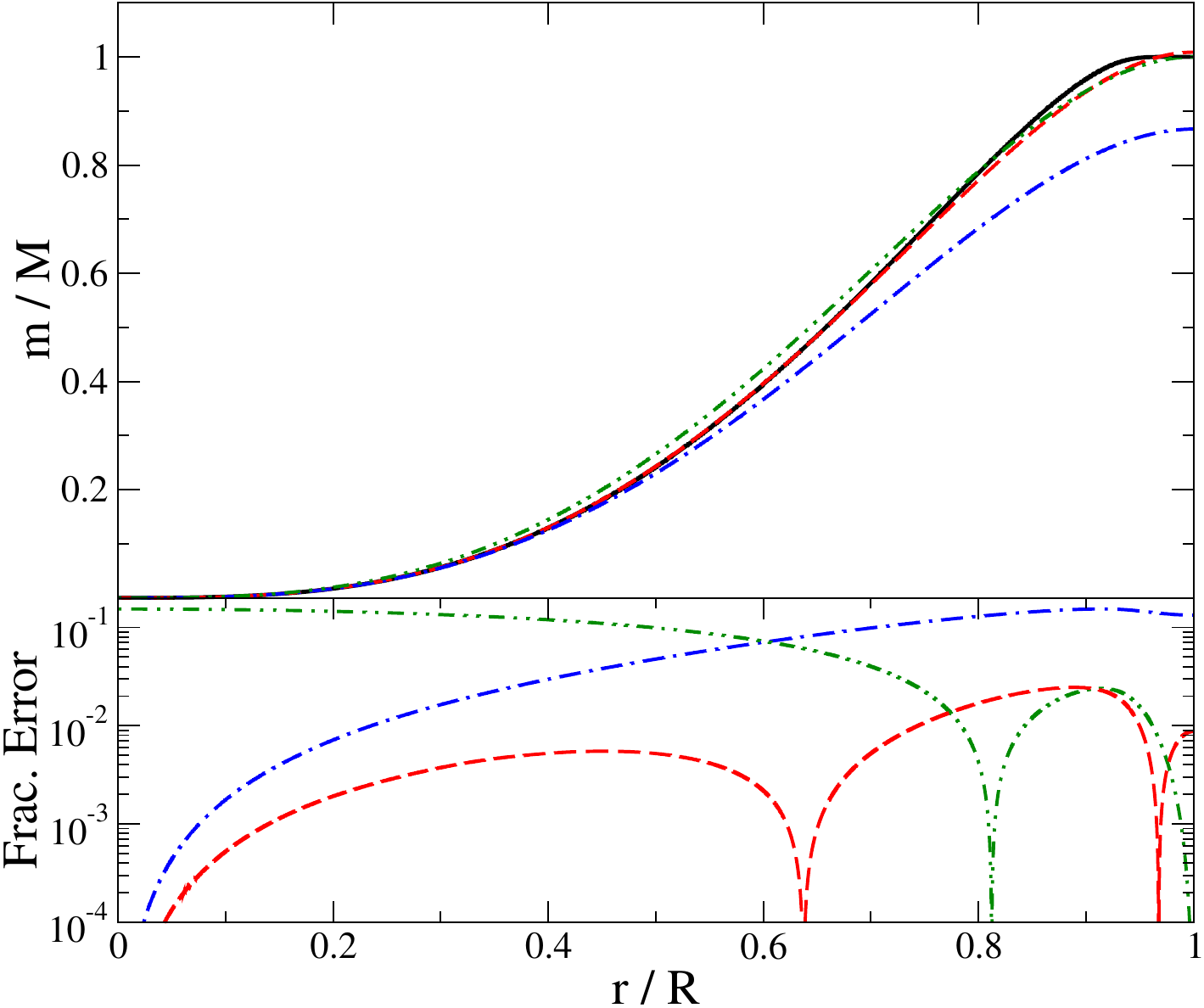}
\includegraphics[width=8.5cm]{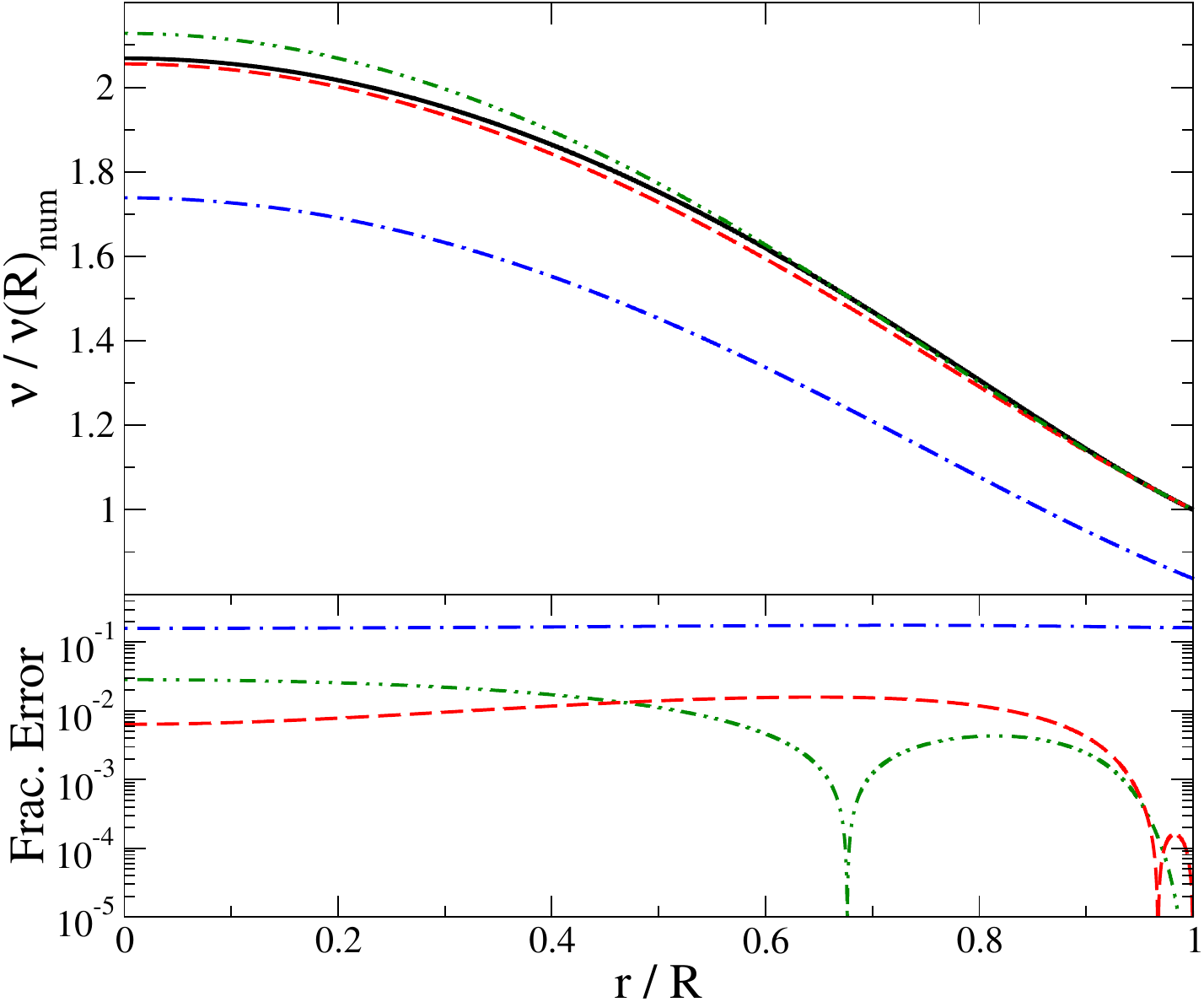}
\includegraphics[width=8.5cm]{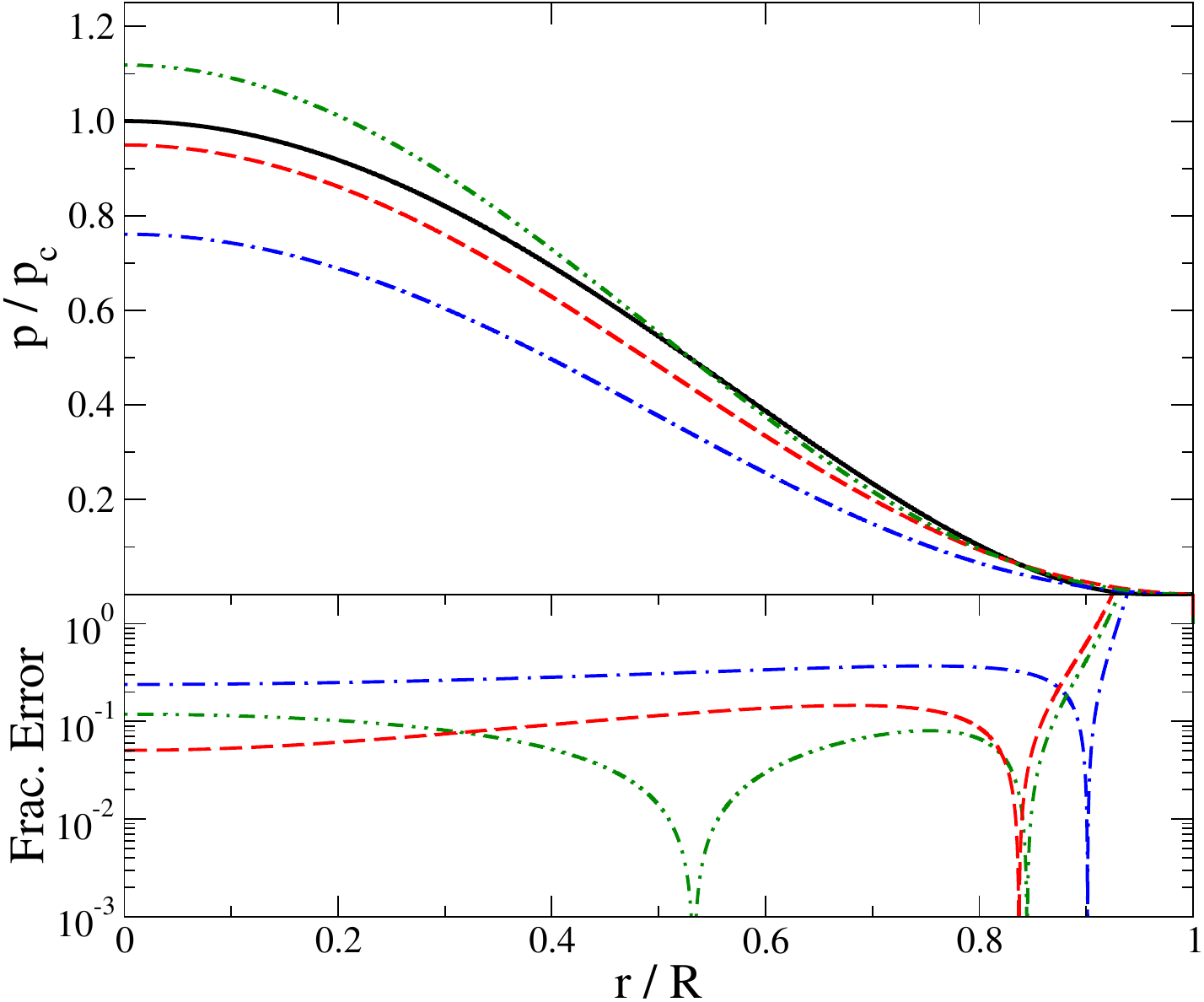}
\caption{(Top) Profiles for energy density $\rho$ using the universal fit for $\alpha$ (top left), interior mass $m$ (top right), $\nu$ (related to the gravitational potential) (bottom left) and pressure $p$ (bottom right) for the original Tolman VII solution in terms of $(R,\rho_c)$ or $(R,M)$ and the improved Tolman VII solution. The values of $\rho_c$, $M$ and $R$ are the same as those in Fig.~\ref{fig:p_rho_Tol}. We also present the numerical result with the AP4 EoS and $M=1.4M_\odot$.
 (Bottom) Fractional errors from the profile obtained numerically.
Observe that the new model works better than the original Tolman solution especially for the $\rho$ and $m$ profiles.
 }
\label{fig:original rho and m profile}
\end{figure*}

\section{Comparison between the Original and Improved Tolman Models}
\label{sec:Comparision}

Let us next compare the original and improved Tolman models against numerical results. We first study the radial profiles of various quantities for a fixed mass and EoS. We then consider root mean square errors (RMSEs) for various masses and EoSs.   

\subsection{Radial Profiles}

We begin by considering radial profiles similar to Fig.~\ref{fig:p_rho_Tol}. Top panels of Fig.~\ref{fig:original rho and m profile} present the $\rho$, $m$, $\nu$ and $p$ profiles of a 1.4$M_\odot $ NS with the AP4 EoS for two different Tolman solutions and the improved model, together with the numerical results. Here, we use the universal fit for $\alpha$.  The bottom panels show the fractional error of each analytic model from the numerical profiles. 

Observe how the new model generally improves the original solution. For example, the $\rho$ and $m$ profiles of the improved model more accurately describe the numerical results over the original Tolman solution. Indeed, the former can fit the realistic profiles within an error of $\sim 5\%$ in most regions of the star. On the other hand, the $\nu$ and $p$ profiles of the improved model are comparable to the original one, though the former is still better than the latter near the stellar center. Both the original and new solutions can model the realistic profiles within an error of $\sim 1\%$ ($\sim 10\%$) for the $\nu$ ($p$) profiles.

\begin{figure*}[htb]
\includegraphics[width=8.5cm]{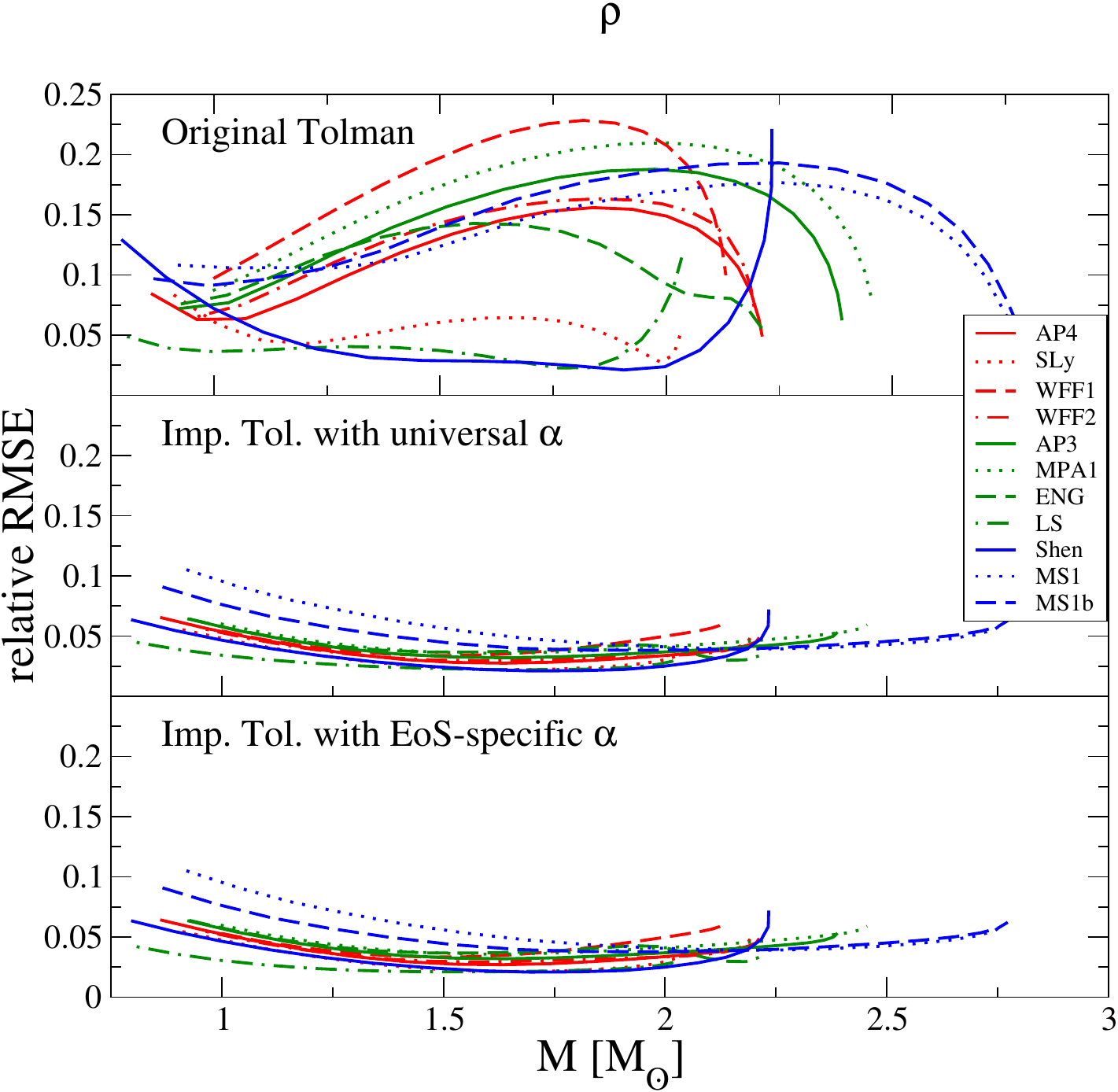}
\includegraphics[width=8.5cm]{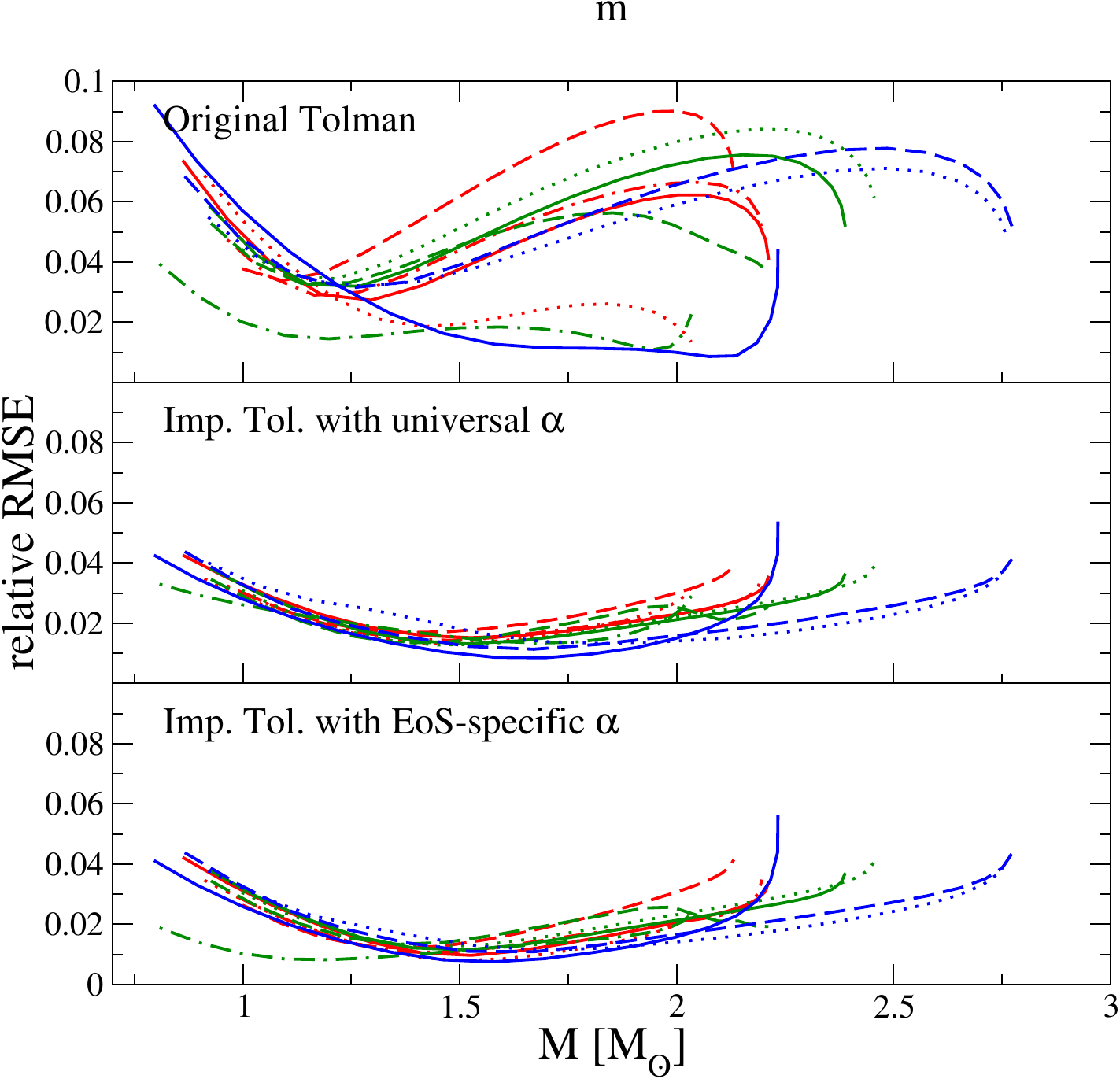}
\includegraphics[width=8.5cm]{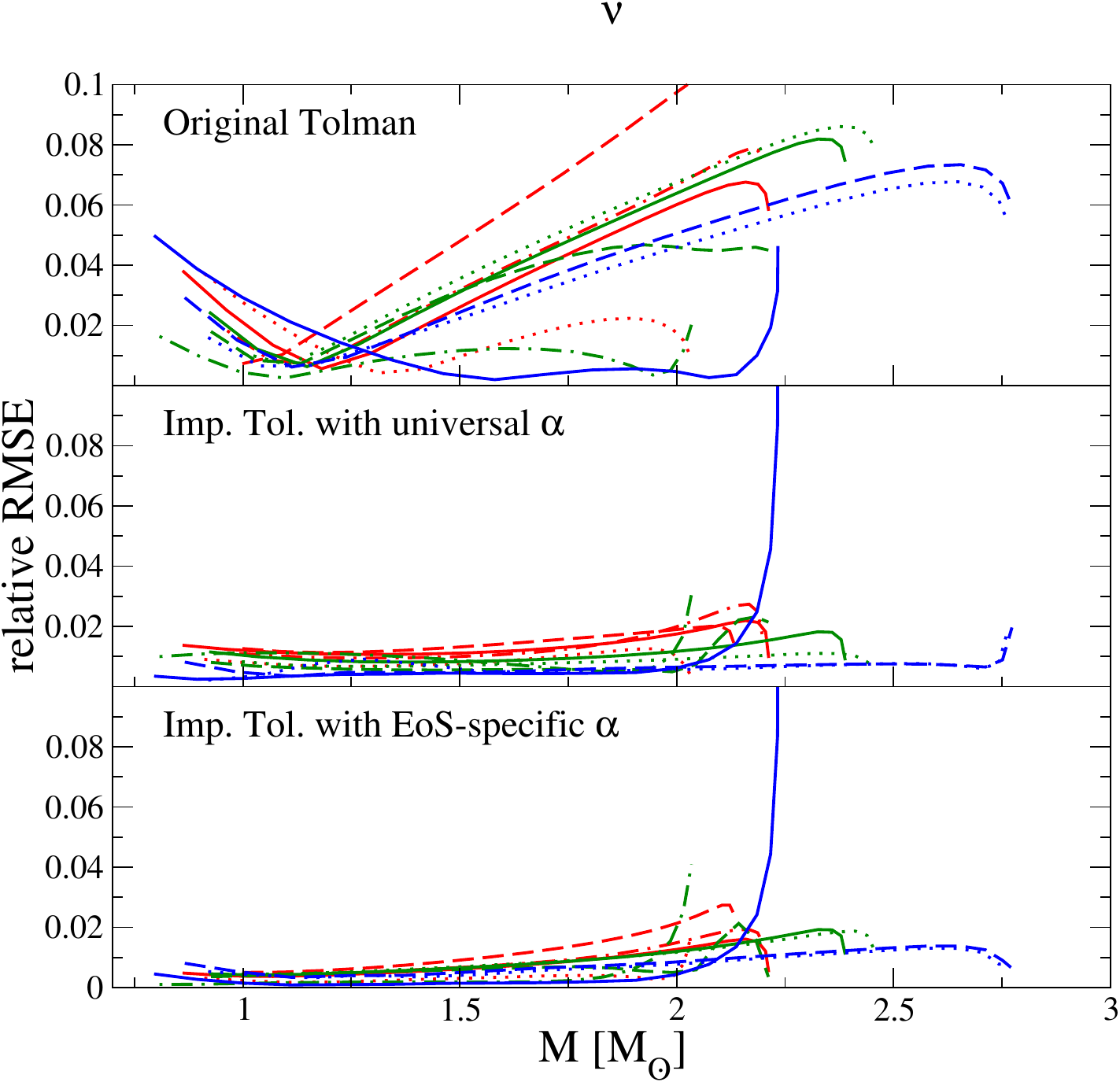}
\includegraphics[width=8.5cm]{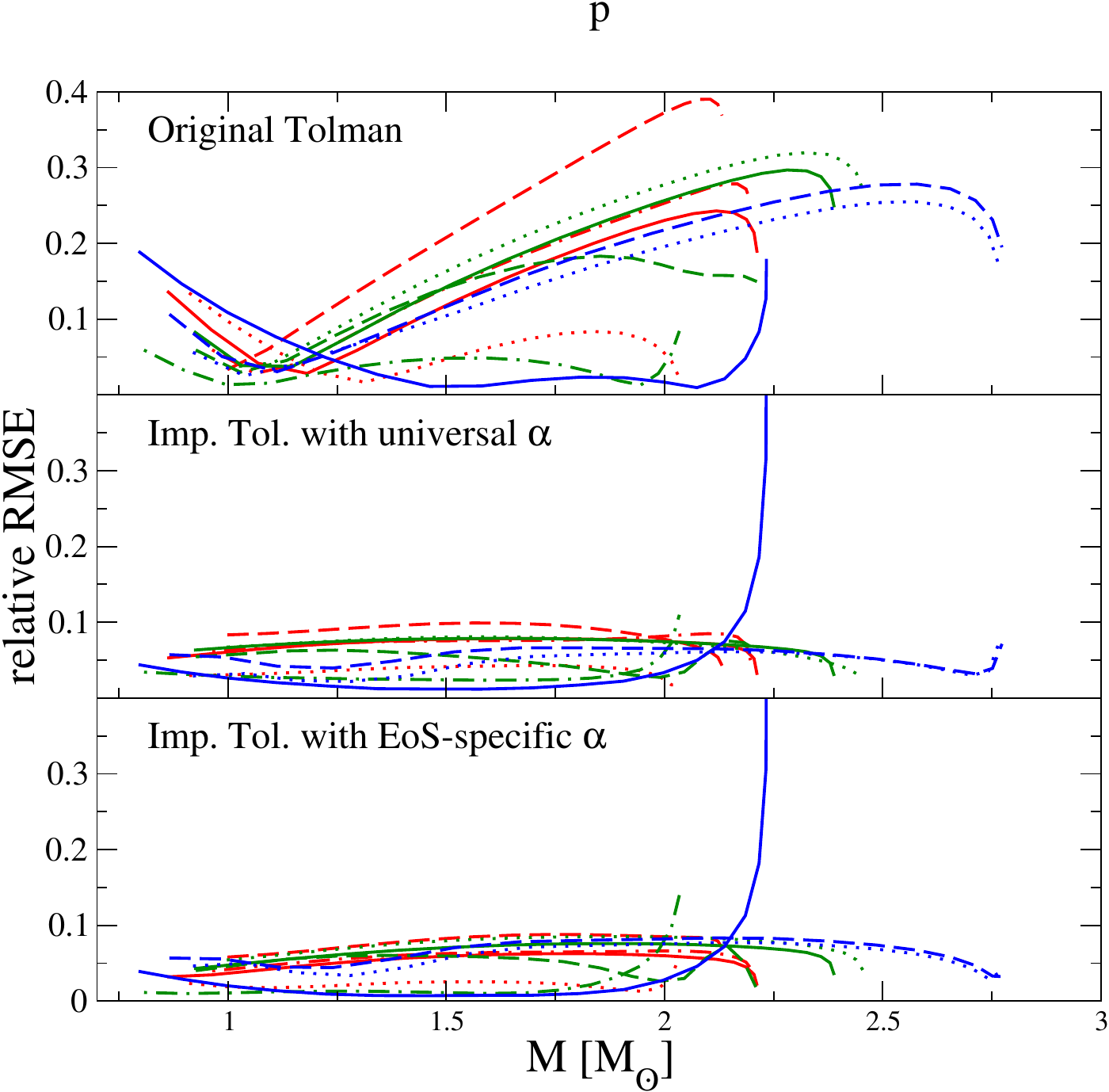}
\caption{The relative RMSE (defined in Eq.~\eqref{eq:RMSE}) of $\rho$ (top left), $m$ (top right), $\nu$ (bottom left) and $p$ (bottom right) for the original Tolman solution with the $(R,M) $ parameterization (top panel),  the improved model with the universal $\alpha$ (middle panel)  and the improved model with the EoS-specific $\alpha$ (bottom panel) as a function of the NS mass, using 11 EoSs with different stiffness in different colors as in Fig.~\ref{fig:alpha}. Observe  how the improved models more accurately describe the realistic profiles (by having smaller relative RMSEs), especially for soft EoSs. }
\label{fig: rmse}
\end{figure*}

\begin{figure*}[htb]
\includegraphics[width=15cm]{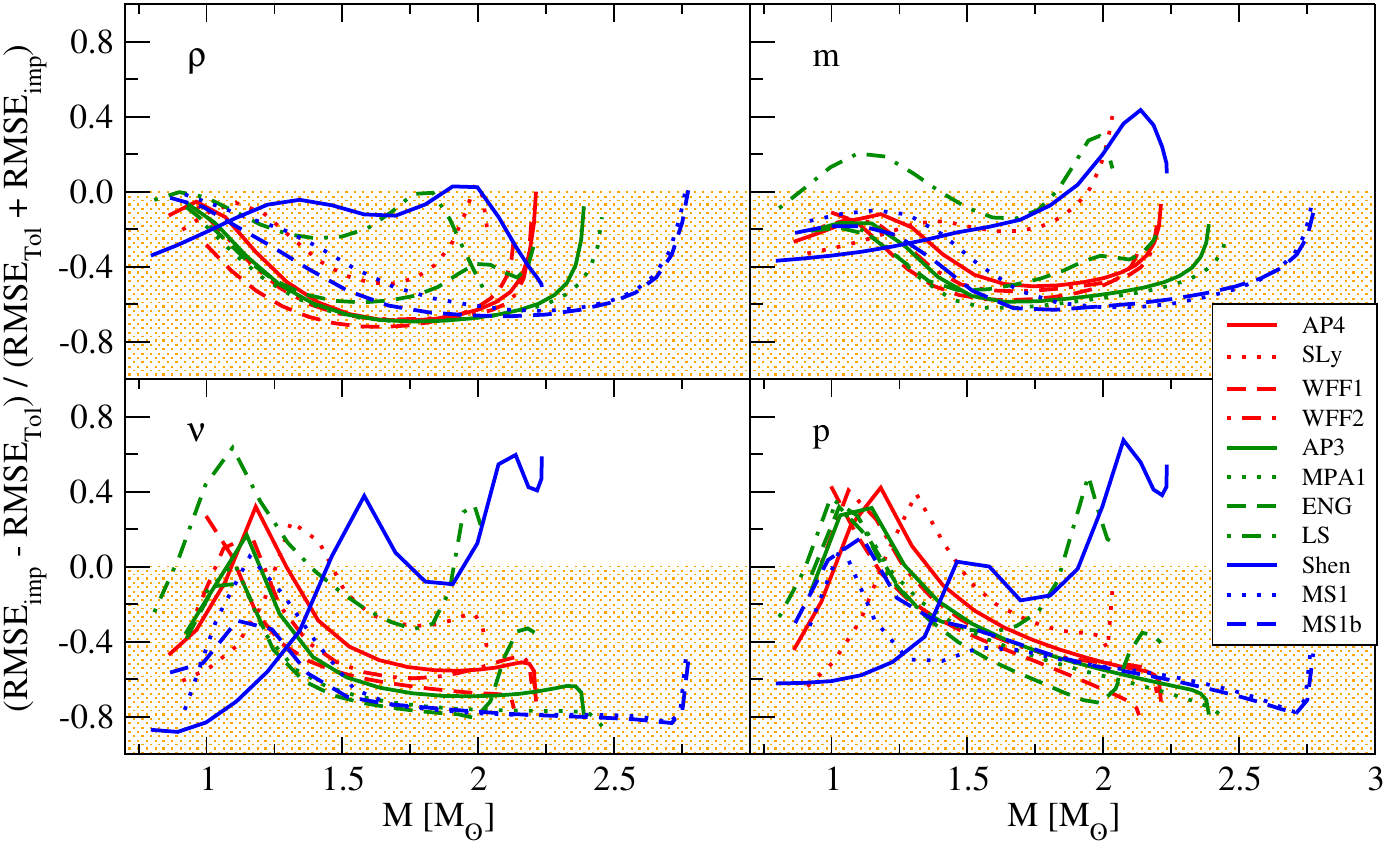}
\caption{The normalized relative RMSE difference between the original Tolman solution and the improved model with the universal $\alpha$. We show the results for $\rho$ (top left), $m$ (top right), $\nu$ (bottom left) and $p$ (bottom right) using the 11 EoSs with different stiffness. The new model has an improvement over the original one if the normalized relative RMSE difference is \emph{negative} (orange shaded region). Observe how the improved model more accurately describes the realistic profiles than the original Tolman solution in most cases.}
\label{fig:URMSE}
\end{figure*}

\begin{figure*}[htb]
\includegraphics[width=15cm]{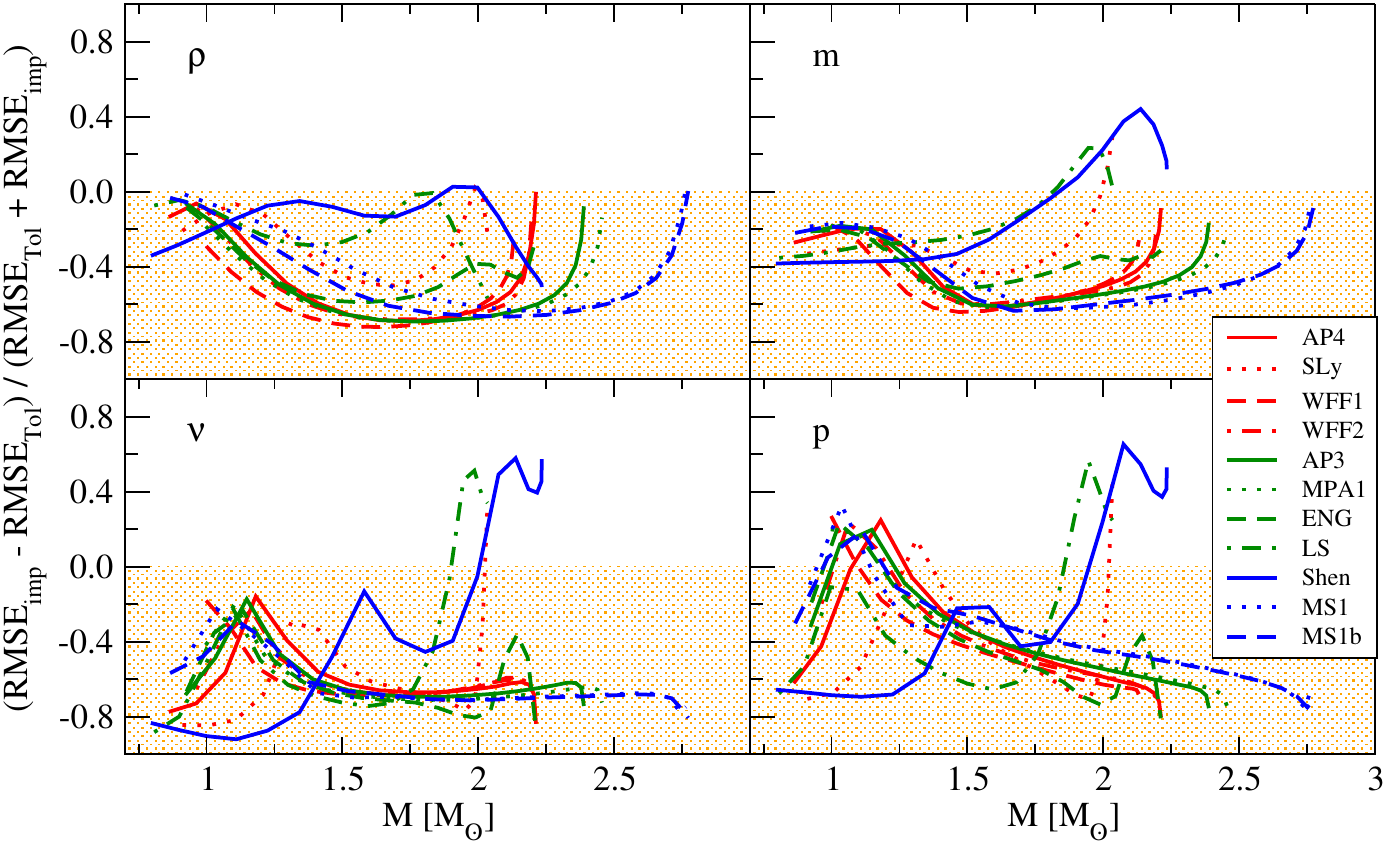}
\caption{
Similar to Fig.~\ref{fig:URMSE} but with the EoS-specific fit for $\alpha$. Observe how the new model further improved from Fig.~\ref{fig:URMSE} by using such EoS-specific fit over the universal one.
}
\label{fig:URMSE2}
\end{figure*}

\subsection{Root Mean Square Errors}

The results presented in the previous subsection was specific to one example NS. How do they change with different masses and EoSs? To address this question, we introduce a relative RMSE, which is a measure of the error of the analytic model from the numerical results throughout the star:
\ba
\label{eq:RMSE}
\mathrm{(relative~RMSE)}=\sqrt{\frac{\int_{0}^{R} \left[y_{\mathrm{num}}(r)-y_{\mathrm{model}}(r)\right]^2 \mathrm{d}r}{\int_{0}^{R} y_{\mathrm{num}}^2(r)\mathrm{d}r}}\,, \nonumber \\
\ea
with $y=(\rho,m,\nu,p)$.

Figure~\ref{fig: rmse} presents the relative RMSE for $\rho$, $m$, $\nu$ and $p$ against the NS mass. We show the relative RMSEs for the 11 EoSs in terms of three models (the original Tolman solution parameterized by $(R,M)$, the improved Tolman models with the universal $\alpha$ and with the EoS-specific $\alpha$).  Observe that in most cases, the improved models have a clear improvement over the original one in terms of accurately describing realistic profiles. This is more significant for soft EoSs (that are more preferred from GW170817), as in the case of the AP4 EoS, where the accuracy improves up to a factor of $\sim 5$ for $\rho$ and $m$. 

To compare the new models against the original one more directly, we show in Fig.~\ref{fig:URMSE} the ratio between the difference and sum of the relative RMSEs for the improved (with the universal $\alpha$) and original Tolman models. The new model more accurately describes numerical profiles than the original one if the ratio is \emph{negative}. Notice first that the energy density profile can be better modeled by the new approximate solution for all EoSs and masses considered here. The situation is similar for the interior mass profile, except for the LS and Shen EoSs. Regarding the gravitational potential ($\nu$) and pressure profiles, the new model performs better especially for soft EoSs with NS masses larger than $1.5M_\odot$.

The accuracy of the new model can be improved further by adopting the EoS-specific fit for $\alpha$, as can be seen from Fig.~\ref{fig:URMSE2}. In this case, the $m$ and $\nu$ profiles for the new model are always better than the original ones, with only exception for high-mass (above $\sim 1.8M_\odot$) NSs with a few EoSs. The $p$ profile has been improved also, though there are some mass ranges (very low mass around 1--1.2$M_\odot$ and very high mass above $1.8M_\odot$) where the original Tolman models performs better. Having said this, the accuracy of the new model is higher than the original one even for the pressure profile in most of the EoSs and the mass range.

\section{Conclusion and Future Directions}
\label{sec:Conclusion}

In this paper, we explore a method to improve the accuracy of the original Tolman VII solution in modeling  numerical solutions. We modified the original expressions by introducing a higher order term in the density profile.  We also succeeded in representing the additionally introduced parameter $\alpha$ in terms of $M$, $R$ and $\rho_c$ in an EoS-insensitive way. The accuracy can be further improved if one uses an EoS-specific fit for $\alpha$. We summarize the expressions for the new model in Table~\ref{table:summary}.

By comparing our results with the numerically solved solutions for 11 different EoSs, we showed that our improved model agrees better with the numerical results than the original Tolman solution. The relative RMSEs for the improved (original) Tolman solution are roughly 10\% (20\%) for energy density, 4\% (10\%) for the interior mass, 2\% (10\%) for the gravitational potential and 10\% (40\%) for pressure. The improvement is significant especially for softer EoSs that are more preferred from GW170817.
 
Future work includes improving the proposed model further. For example, one may come up with a more appropriate density profile that can correctly capture its behavior close to the stellar surface. One can also try to find different ways of finding approximate solutions to the Einstein equations that will improve the modeling. The model presented here does not apply to stellar solutions whose density does not vanish at the surface, such as quark stars and self-bound stars~\cite{Chan:2015iou}. It would be interesting to construct analytic interior models appropriate for these kinds of stars. One could also try to improve other analytic solutions, such as the one found by Buchdahl~\cite{1967ApJ...147..310B,Schutz:1985jx} which was compared against realistic neutron star solutions in~\cite{0004-637X-550-1-426}.  

Yet, another possible avenue is to extend the analysis presented here to more realistic NSs with rotation or tidal deformation. The first thing one can try is to assume these effects are small and treat them as perturbation to the solution presented here. If one can construct such solutions analytically, one can extract global quantities like the stellar moment of inertia, tidal Love number and quadrupole moment, among which universal I-Love-Q relations are known to exist~\cite{I-Love-Q-Science,I-Love-Q-PRD,Yagi:2016bkt,Doneva:2017jop}. Such analytic study may help understand the origin of the universality. Work along this line is currently in progress.

\acknowledgments
KY acknowledges support from NSF Award PHY-1806776. 
K.Y. would like to also acknowledge networking support by the COST Action GWverse CA16104.
%



\newpage

\bibliography{Reference}

\begin{thebibliography}{47}%
\makeatletter
\providecommand \@ifxundefined [1]{%
 \@ifx{#1\undefined}
}%
\providecommand \@ifnum [1]{%
 \ifnum #1\expandafter \@firstoftwo
 \else \expandafter \@secondoftwo
 \fi
}%
\providecommand \@ifx [1]{%
 \ifx #1\expandafter \@firstoftwo
 \else \expandafter \@secondoftwo
 \fi
}%
\providecommand \natexlab [1]{#1}%
\providecommand \enquote  [1]{``#1''}%
\providecommand \bibnamefont  [1]{#1}%
\providecommand \bibfnamefont [1]{#1}%
\providecommand \citenamefont [1]{#1}%
\providecommand \href@noop [0]{\@secondoftwo}%
\providecommand \href [0]{\begingroup \@sanitize@url \@href}%
\providecommand \@href[1]{\@@startlink{#1}\@@href}%
\providecommand \@@href[1]{\endgroup#1\@@endlink}%
\providecommand \@sanitize@url [0]{\catcode `\\12\catcode `\$12\catcode
  `\&12\catcode `\#12\catcode `\^12\catcode `\_12\catcode `\%12\relax}%
\providecommand \@@startlink[1]{}%
\providecommand \@@endlink[0]{}%
\providecommand \url  [0]{\begingroup\@sanitize@url \@url }%
\providecommand \@url [1]{\endgroup\@href {#1}{\urlprefix }}%
\providecommand \urlprefix  [0]{URL }%
\providecommand \Eprint [0]{\href }%
\providecommand \doibase [0]{http://dx.doi.org/}%
\providecommand \selectlanguage [0]{\@gobble}%
\providecommand \bibinfo  [0]{\@secondoftwo}%
\providecommand \bibfield  [0]{\@secondoftwo}%
\providecommand \translation [1]{[#1]}%
\providecommand \BibitemOpen [0]{}%
\providecommand \bibitemStop [0]{}%
\providecommand \bibitemNoStop [0]{.\EOS\space}%
\providecommand \EOS [0]{\spacefactor3000\relax}%
\providecommand \BibitemShut  [1]{\csname bibitem#1\endcsname}%
\let\auto@bib@innerbib\@empty
\bibitem [{\citenamefont {Lattimer}\ and\ \citenamefont
  {Prakash}(2001)}]{0004-637X-550-1-426}%
  \BibitemOpen
  \bibfield  {author} {\bibinfo {author} {\bibfnamefont {J.~M.}\ \bibnamefont
  {Lattimer}}\ and\ \bibinfo {author} {\bibfnamefont {M.}~\bibnamefont
  {Prakash}},\ }\href {http://stacks.iop.org/0004-637X/550/i=1/a=426}
  {\bibfield  {journal} {\bibinfo  {journal} {The Astrophysical Journal}\
  }\textbf {\bibinfo {volume} {550}},\ \bibinfo {pages} {426} (\bibinfo {year}
  {2001})}\BibitemShut {NoStop}%
\bibitem [{\citenamefont {Lattimer}\ and\ \citenamefont
  {Prakash}(2007)}]{lattimer-prakash-review}%
  \BibitemOpen
  \bibfield  {author} {\bibinfo {author} {\bibfnamefont {J.~M.}\ \bibnamefont
  {Lattimer}}\ and\ \bibinfo {author} {\bibfnamefont {M.}~\bibnamefont
  {Prakash}},\ }\href {\doibase 10.1016/j.physrep.2007.02.003} {\bibfield
  {journal} {\bibinfo  {journal} {Phys.Rept.}\ }\textbf {\bibinfo {volume}
  {442}},\ \bibinfo {pages} {109} (\bibinfo {year} {2007})}\BibitemShut
  {NoStop}%
\bibitem [{\citenamefont {Özel}\ and\ \citenamefont
  {Freire}(2016)}]{Ozel:2016oaf}%
  \BibitemOpen
  \bibfield  {author} {\bibinfo {author} {\bibfnamefont {F.}~\bibnamefont
  {Özel}}\ and\ \bibinfo {author} {\bibfnamefont {P.}~\bibnamefont {Freire}},\
  }\href {\doibase 10.1146/annurev-astro-081915-023322} {\bibfield  {journal}
  {\bibinfo  {journal} {Ann. Rev. Astron. Astrophys.}\ }\textbf {\bibinfo
  {volume} {54}},\ \bibinfo {pages} {401} (\bibinfo {year} {2016})},\ \Eprint
  {http://arxiv.org/abs/1603.02698} {arXiv:1603.02698 [astro-ph.HE]}
  \BibitemShut {NoStop}%
\bibitem [{\citenamefont {Ozel}\ \emph {et~al.}(2010)\citenamefont {Ozel},
  \citenamefont {Baym},\ and\ \citenamefont {Guver}}]{ozel-baym-guver}%
  \BibitemOpen
  \bibfield  {author} {\bibinfo {author} {\bibfnamefont {F.}~\bibnamefont
  {Ozel}}, \bibinfo {author} {\bibfnamefont {G.}~\bibnamefont {Baym}}, \ and\
  \bibinfo {author} {\bibfnamefont {T.}~\bibnamefont {Guver}},\ }\href
  {\doibase 10.1103/PhysRevD.82.101301} {\bibfield  {journal} {\bibinfo
  {journal} {Phys.Rev.}\ }\textbf {\bibinfo {volume} {D82}},\ \bibinfo {pages}
  {101301} (\bibinfo {year} {2010})},\ \Eprint {http://arxiv.org/abs/1002.3153}
  {arXiv:1002.3153 [astro-ph.HE]} \BibitemShut {NoStop}%
\bibitem [{\citenamefont {Steiner}\ \emph {et~al.}(2010)\citenamefont
  {Steiner}, \citenamefont {Lattimer},\ and\ \citenamefont
  {Brown}}]{steiner-lattimer-brown}%
  \BibitemOpen
  \bibfield  {author} {\bibinfo {author} {\bibfnamefont {A.~W.}\ \bibnamefont
  {Steiner}}, \bibinfo {author} {\bibfnamefont {J.~M.}\ \bibnamefont
  {Lattimer}}, \ and\ \bibinfo {author} {\bibfnamefont {E.~F.}\ \bibnamefont
  {Brown}},\ }\href {\doibase 10.1088/0004-637X/722/1/33} {\bibfield  {journal}
  {\bibinfo  {journal} {Astrophys.J.}\ }\textbf {\bibinfo {volume} {722}},\
  \bibinfo {pages} {33} (\bibinfo {year} {2010})}\BibitemShut {NoStop}%
\bibitem [{\citenamefont {Ozel}\ \emph
  {et~al.}(2016{\natexlab{a}})\citenamefont {Ozel}, \citenamefont {Psaltis},
  \citenamefont {Guver}, \citenamefont {Baym}, \citenamefont {Heinke},\ and\
  \citenamefont {Guillot}}]{Ozel:2015fia}%
  \BibitemOpen
  \bibfield  {author} {\bibinfo {author} {\bibfnamefont {F.}~\bibnamefont
  {Ozel}}, \bibinfo {author} {\bibfnamefont {D.}~\bibnamefont {Psaltis}},
  \bibinfo {author} {\bibfnamefont {T.}~\bibnamefont {Guver}}, \bibinfo
  {author} {\bibfnamefont {G.}~\bibnamefont {Baym}}, \bibinfo {author}
  {\bibfnamefont {C.}~\bibnamefont {Heinke}}, \ and\ \bibinfo {author}
  {\bibfnamefont {S.}~\bibnamefont {Guillot}},\ }\href {\doibase
  10.3847/0004-637X/820/1/28} {\bibfield  {journal} {\bibinfo  {journal}
  {Astrophys. J.}\ }\textbf {\bibinfo {volume} {820}},\ \bibinfo {pages} {28}
  (\bibinfo {year} {2016}{\natexlab{a}})},\ \Eprint
  {http://arxiv.org/abs/1505.05155} {arXiv:1505.05155 [astro-ph.HE]}
  \BibitemShut {NoStop}%
\bibitem [{\citenamefont {Ozel}\ \emph
  {et~al.}(2016{\natexlab{b}})\citenamefont {Ozel}, \citenamefont {Psaltis},
  \citenamefont {Arzoumanian}, \citenamefont {Morsink},\ and\ \citenamefont
  {Baubock}}]{Ozel:2015ykl}%
  \BibitemOpen
  \bibfield  {author} {\bibinfo {author} {\bibfnamefont {F.}~\bibnamefont
  {Ozel}}, \bibinfo {author} {\bibfnamefont {D.}~\bibnamefont {Psaltis}},
  \bibinfo {author} {\bibfnamefont {Z.}~\bibnamefont {Arzoumanian}}, \bibinfo
  {author} {\bibfnamefont {S.}~\bibnamefont {Morsink}}, \ and\ \bibinfo
  {author} {\bibfnamefont {M.}~\bibnamefont {Baubock}},\ }\href {\doibase
  10.3847/0004-637X/832/1/92} {\bibfield  {journal} {\bibinfo  {journal}
  {Astrophys. J.}\ }\textbf {\bibinfo {volume} {832}},\ \bibinfo {pages} {92}
  (\bibinfo {year} {2016}{\natexlab{b}})},\ \Eprint
  {http://arxiv.org/abs/1512.03067} {arXiv:1512.03067 [astro-ph.HE]}
  \BibitemShut {NoStop}%
\bibitem [{\citenamefont {Lo}\ \emph {et~al.}(2013)\citenamefont {Lo},
  \citenamefont {Coleman~Miller}, \citenamefont {Bhattacharyya},\ and\
  \citenamefont {Lamb}}]{Lo:2013ava}%
  \BibitemOpen
  \bibfield  {author} {\bibinfo {author} {\bibfnamefont {K.~H.}\ \bibnamefont
  {Lo}}, \bibinfo {author} {\bibfnamefont {M.}~\bibnamefont {Coleman~Miller}},
  \bibinfo {author} {\bibfnamefont {S.}~\bibnamefont {Bhattacharyya}}, \ and\
  \bibinfo {author} {\bibfnamefont {F.~K.}\ \bibnamefont {Lamb}},\ }\href
  {\doibase 10.1088/0004-637X/776/1/19} {\bibfield  {journal} {\bibinfo
  {journal} {Astrophys.J.}\ }\textbf {\bibinfo {volume} {776}},\ \bibinfo
  {pages} {19} (\bibinfo {year} {2013})},\ \Eprint
  {http://arxiv.org/abs/1304.2330} {arXiv:1304.2330 [astro-ph.HE]} \BibitemShut
  {NoStop}%
\bibitem [{\citenamefont {Lo}\ \emph {et~al.}(2018)\citenamefont {Lo},
  \citenamefont {Miller}, \citenamefont {Bhattacharyya},\ and\ \citenamefont
  {Lamb}}]{Lo:2018hes}%
  \BibitemOpen
  \bibfield  {author} {\bibinfo {author} {\bibfnamefont {K.~H.}\ \bibnamefont
  {Lo}}, \bibinfo {author} {\bibfnamefont {M.~C.}\ \bibnamefont {Miller}},
  \bibinfo {author} {\bibfnamefont {S.}~\bibnamefont {Bhattacharyya}}, \ and\
  \bibinfo {author} {\bibfnamefont {F.~K.}\ \bibnamefont {Lamb}},\ }\href
  {\doibase 10.3847/1538-4357/aaa95b} {\bibfield  {journal} {\bibinfo
  {journal} {Astrophys. J.}\ }\textbf {\bibinfo {volume} {854}},\ \bibinfo
  {pages} {187} (\bibinfo {year} {2018})},\ \Eprint
  {http://arxiv.org/abs/1801.08031} {arXiv:1801.08031 [astro-ph.HE]}
  \BibitemShut {NoStop}%
\bibitem [{\citenamefont {Abbott}\ \emph
  {et~al.}(2017{\natexlab{a}})\citenamefont {Abbott} \emph
  {et~al.}}]{PhysRevLett.119.161101}%
  \BibitemOpen
  \bibfield  {author} {\bibinfo {author} {\bibfnamefont {B.~P.}\ \bibnamefont
  {Abbott}} \emph {et~al.} (\bibinfo {collaboration} {Virgo, LIGO
  Scientific}),\ }\href {\doibase 10.1103/PhysRevLett.119.161101} {\bibfield
  {journal} {\bibinfo  {journal} {Phys. Rev. Lett.}\ }\textbf {\bibinfo
  {volume} {119}},\ \bibinfo {pages} {161101} (\bibinfo {year}
  {2017}{\natexlab{a}})},\ \Eprint {http://arxiv.org/abs/1710.05832}
  {arXiv:1710.05832 [gr-qc]} \BibitemShut {NoStop}%
\bibitem [{\citenamefont {Abbott}\ \emph {et~al.}(2019)\citenamefont {Abbott}
  \emph {et~al.}}]{Abbott:2018wiz}%
  \BibitemOpen
  \bibfield  {author} {\bibinfo {author} {\bibfnamefont {B.~P.}\ \bibnamefont
  {Abbott}} \emph {et~al.} (\bibinfo {collaboration} {LIGO Scientific,
  Virgo}),\ }\href {\doibase 10.1103/PhysRevX.9.011001} {\bibfield  {journal}
  {\bibinfo  {journal} {Phys. Rev.}\ }\textbf {\bibinfo {volume} {X9}},\
  \bibinfo {pages} {011001} (\bibinfo {year} {2019})},\ \Eprint
  {http://arxiv.org/abs/1805.11579} {arXiv:1805.11579 [gr-qc]} \BibitemShut
  {NoStop}%
\bibitem [{\citenamefont {Abbott}\ \emph
  {et~al.}(2018{\natexlab{a}})\citenamefont {Abbott} \emph
  {et~al.}}]{Abbott:2018exr}%
  \BibitemOpen
  \bibfield  {author} {\bibinfo {author} {\bibfnamefont {B.~P.}\ \bibnamefont
  {Abbott}} \emph {et~al.} (\bibinfo {collaboration} {LIGO Scientific,
  Virgo}),\ }\href {\doibase 10.1103/PhysRevLett.121.161101} {\bibfield
  {journal} {\bibinfo  {journal} {Phys. Rev. Lett.}\ }\textbf {\bibinfo
  {volume} {121}},\ \bibinfo {pages} {161101} (\bibinfo {year}
  {2018}{\natexlab{a}})},\ \Eprint {http://arxiv.org/abs/1805.11581}
  {arXiv:1805.11581 [gr-qc]} \BibitemShut {NoStop}%
\bibitem [{\citenamefont {Malik}\ \emph {et~al.}(2018)\citenamefont {Malik},
  \citenamefont {Alam}, \citenamefont {Fortin}, \citenamefont {Providência},
  \citenamefont {Agrawal}, \citenamefont {Jha}, \citenamefont {Kumar},\ and\
  \citenamefont {Patra}}]{Malik:2018zcf}%
  \BibitemOpen
  \bibfield  {author} {\bibinfo {author} {\bibfnamefont {T.}~\bibnamefont
  {Malik}}, \bibinfo {author} {\bibfnamefont {N.}~\bibnamefont {Alam}},
  \bibinfo {author} {\bibfnamefont {M.}~\bibnamefont {Fortin}}, \bibinfo
  {author} {\bibfnamefont {C.}~\bibnamefont {Providência}}, \bibinfo {author}
  {\bibfnamefont {B.~K.}\ \bibnamefont {Agrawal}}, \bibinfo {author}
  {\bibfnamefont {T.~K.}\ \bibnamefont {Jha}}, \bibinfo {author} {\bibfnamefont
  {B.}~\bibnamefont {Kumar}}, \ and\ \bibinfo {author} {\bibfnamefont {S.~K.}\
  \bibnamefont {Patra}},\ }\href {\doibase 10.1103/PhysRevC.98.035804}
  {\bibfield  {journal} {\bibinfo  {journal} {Phys. Rev.}\ }\textbf {\bibinfo
  {volume} {C98}},\ \bibinfo {pages} {035804} (\bibinfo {year} {2018})},\
  \Eprint {http://arxiv.org/abs/1805.11963} {arXiv:1805.11963 [nucl-th]}
  \BibitemShut {NoStop}%
\bibitem [{\citenamefont {Carson}\ \emph {et~al.}(2018)\citenamefont {Carson},
  \citenamefont {Steiner},\ and\ \citenamefont {Yagi}}]{Carson:2018xri}%
  \BibitemOpen
  \bibfield  {author} {\bibinfo {author} {\bibfnamefont {Z.}~\bibnamefont
  {Carson}}, \bibinfo {author} {\bibfnamefont {A.~W.}\ \bibnamefont {Steiner}},
  \ and\ \bibinfo {author} {\bibfnamefont {K.}~\bibnamefont {Yagi}},\
  }\href@noop {} {\  (\bibinfo {year} {2018})},\ \Eprint
  {http://arxiv.org/abs/1812.08910} {arXiv:1812.08910 [gr-qc]} \BibitemShut
  {NoStop}%
\bibitem [{\citenamefont {Kramer}\ \emph {et~al.}(2006)\citenamefont {Kramer},
  \citenamefont {Stairs}, \citenamefont {Manchester}, \citenamefont
  {McLaughlin}, \citenamefont {Lyne}, \citenamefont {Ferdman}, \citenamefont
  {Burgay}, \citenamefont {Lorimer}, \citenamefont {Possenti}, \citenamefont
  {D{\textquoteright}Amico}, \citenamefont {Sarkissian}, \citenamefont {Hobbs},
  \citenamefont {Reynolds}, \citenamefont {Freire},\ and\ \citenamefont
  {Camilo}}]{GRtest_pulsar}%
  \BibitemOpen
  \bibfield  {author} {\bibinfo {author} {\bibfnamefont {M.}~\bibnamefont
  {Kramer}}, \bibinfo {author} {\bibfnamefont {I.~H.}\ \bibnamefont {Stairs}},
  \bibinfo {author} {\bibfnamefont {R.~N.}\ \bibnamefont {Manchester}},
  \bibinfo {author} {\bibfnamefont {M.~A.}\ \bibnamefont {McLaughlin}},
  \bibinfo {author} {\bibfnamefont {A.~G.}\ \bibnamefont {Lyne}}, \bibinfo
  {author} {\bibfnamefont {R.~D.}\ \bibnamefont {Ferdman}}, \bibinfo {author}
  {\bibfnamefont {M.}~\bibnamefont {Burgay}}, \bibinfo {author} {\bibfnamefont
  {D.~R.}\ \bibnamefont {Lorimer}}, \bibinfo {author} {\bibfnamefont
  {A.}~\bibnamefont {Possenti}}, \bibinfo {author} {\bibfnamefont
  {N.}~\bibnamefont {D{\textquoteright}Amico}}, \bibinfo {author}
  {\bibfnamefont {J.~M.}\ \bibnamefont {Sarkissian}}, \bibinfo {author}
  {\bibfnamefont {G.~B.}\ \bibnamefont {Hobbs}}, \bibinfo {author}
  {\bibfnamefont {J.~E.}\ \bibnamefont {Reynolds}}, \bibinfo {author}
  {\bibfnamefont {P.~C.~C.}\ \bibnamefont {Freire}}, \ and\ \bibinfo {author}
  {\bibfnamefont {F.}~\bibnamefont {Camilo}},\ }\href {\doibase
  10.1126/science.1132305} {\bibfield  {journal} {\bibinfo  {journal}
  {Science}\ }\textbf {\bibinfo {volume} {314}},\ \bibinfo {pages} {97}
  (\bibinfo {year} {2006})},\ \Eprint
  {http://arxiv.org/abs/http://science.sciencemag.org/content/314/5796/97.full.pdf}
  {http://science.sciencemag.org/content/314/5796/97.full.pdf} \BibitemShut
  {NoStop}%
\bibitem [{\citenamefont {Stairs}(2003)}]{stairs}%
  \BibitemOpen
  \bibfield  {author} {\bibinfo {author} {\bibfnamefont {I.~H.}\ \bibnamefont
  {Stairs}},\ }\href@noop {} {\bibfield  {journal} {\bibinfo  {journal} {Living
  Rev.Rel.}\ }\textbf {\bibinfo {volume} {6}},\ \bibinfo {pages} {5} (\bibinfo
  {year} {2003})}\BibitemShut {NoStop}%
\bibitem [{\citenamefont {Abbott}\ \emph
  {et~al.}(2017{\natexlab{b}})\citenamefont {Abbott} \emph
  {et~al.}}]{Monitor:2017mdv}%
  \BibitemOpen
  \bibfield  {author} {\bibinfo {author} {\bibfnamefont {B.~P.}\ \bibnamefont
  {Abbott}} \emph {et~al.} (\bibinfo {collaboration} {LIGO Scientific, Virgo,
  Fermi-GBM, INTEGRAL}),\ }\href {\doibase 10.3847/2041-8213/aa920c} {\bibfield
   {journal} {\bibinfo  {journal} {Astrophys. J.}\ }\textbf {\bibinfo {volume}
  {848}},\ \bibinfo {pages} {L13} (\bibinfo {year} {2017}{\natexlab{b}})},\
  \Eprint {http://arxiv.org/abs/1710.05834} {arXiv:1710.05834 [astro-ph.HE]}
  \BibitemShut {NoStop}%
\bibitem [{\citenamefont {Abbott}\ \emph
  {et~al.}(2018{\natexlab{b}})\citenamefont {Abbott} \emph
  {et~al.}}]{Abbott:2018lct}%
  \BibitemOpen
  \bibfield  {author} {\bibinfo {author} {\bibfnamefont {B.~P.}\ \bibnamefont
  {Abbott}} \emph {et~al.} (\bibinfo {collaboration} {LIGO Scientific,
  Virgo}),\ }\href@noop {} {\  (\bibinfo {year} {2018}{\natexlab{b}})},\
  \Eprint {http://arxiv.org/abs/1811.00364} {arXiv:1811.00364 [gr-qc]}
  \BibitemShut {NoStop}%
\bibitem [{\citenamefont {Schutz}(1985)}]{Schutz:1985jx}%
  \BibitemOpen
  \bibfield  {author} {\bibinfo {author} {\bibfnamefont {B.~F.}\ \bibnamefont
  {Schutz}},\ }\href@noop {} {\emph {\bibinfo {title} {{A FIRST COURSE IN
  GENERAL RELATIVITY}}}}\ (\bibinfo  {publisher} {Cambridge Univ. Pr.},\
  \bibinfo {address} {Cambridge, UK},\ \bibinfo {year} {1985})\BibitemShut
  {NoStop}%
\bibitem [{\citenamefont {{Buchdahl}}(1967)}]{1967ApJ...147..310B}%
  \BibitemOpen
  \bibfield  {author} {\bibinfo {author} {\bibfnamefont {H.~A.}\ \bibnamefont
  {{Buchdahl}}},\ }\href {\doibase 10.1086/149001} {\bibfield  {journal}
  {\bibinfo  {journal} {\apj}\ }\textbf {\bibinfo {volume} {147}},\ \bibinfo
  {pages} {310} (\bibinfo {year} {1967})}\BibitemShut {NoStop}%
\bibitem [{\citenamefont {Tolman}(1939)}]{PhysRev.55.364}%
  \BibitemOpen
  \bibfield  {author} {\bibinfo {author} {\bibfnamefont {R.~C.}\ \bibnamefont
  {Tolman}},\ }\href {\doibase 10.1103/PhysRev.55.364} {\bibfield  {journal}
  {\bibinfo  {journal} {Phys. Rev.}\ }\textbf {\bibinfo {volume} {55}},\
  \bibinfo {pages} {364} (\bibinfo {year} {1939})}\BibitemShut {NoStop}%
\bibitem [{\citenamefont {Raghoonundun}\ and\ \citenamefont
  {Hobill}(2015)}]{Raghoonundun:2015wga}%
  \BibitemOpen
  \bibfield  {author} {\bibinfo {author} {\bibfnamefont {A.~M.}\ \bibnamefont
  {Raghoonundun}}\ and\ \bibinfo {author} {\bibfnamefont {D.~W.}\ \bibnamefont
  {Hobill}},\ }\href {\doibase 10.1103/PhysRevD.92.124005} {\bibfield
  {journal} {\bibinfo  {journal} {Phys. Rev.}\ }\textbf {\bibinfo {volume}
  {D92}},\ \bibinfo {pages} {124005} (\bibinfo {year} {2015})},\ \Eprint
  {http://arxiv.org/abs/1506.05813} {arXiv:1506.05813 [gr-qc]} \BibitemShut
  {NoStop}%
\bibitem [{\citenamefont {P.S.~Negi}(2001)}]{TolmanVIIstability}%
  \BibitemOpen
  \bibfield  {author} {\bibinfo {author} {\bibfnamefont {M.~D.}\ \bibnamefont
  {P.S.~Negi}},\ }\href {\doibase /10.1023/A:1002707730439} {\bibfield
  {journal} {\bibinfo  {journal} {Astrophysics and Space Science}\ }\textbf
  {\bibinfo {volume} {275}},\ \bibinfo {pages} {185} (\bibinfo {year}
  {2001})}\BibitemShut {NoStop}%
\bibitem [{\citenamefont {Neary}\ \emph {et~al.}(2001)\citenamefont {Neary},
  \citenamefont {Ishak},\ and\ \citenamefont {Lake}}]{Neary:2001ai}%
  \BibitemOpen
  \bibfield  {author} {\bibinfo {author} {\bibfnamefont {N.}~\bibnamefont
  {Neary}}, \bibinfo {author} {\bibfnamefont {M.}~\bibnamefont {Ishak}}, \ and\
  \bibinfo {author} {\bibfnamefont {K.}~\bibnamefont {Lake}},\ }\href {\doibase
  10.1103/PhysRevD.64.084001} {\bibfield  {journal} {\bibinfo  {journal} {Phys.
  Rev.}\ }\textbf {\bibinfo {volume} {D64}},\ \bibinfo {pages} {084001}
  (\bibinfo {year} {2001})},\ \Eprint {http://arxiv.org/abs/gr-qc/0104002}
  {arXiv:gr-qc/0104002 [gr-qc]} \BibitemShut {NoStop}%
\bibitem [{\citenamefont {Raghoonundun}\ and\ \citenamefont
  {Hobill}(2016)}]{Raghoonundun:2016cun}%
  \BibitemOpen
  \bibfield  {author} {\bibinfo {author} {\bibfnamefont {A.~M.}\ \bibnamefont
  {Raghoonundun}}\ and\ \bibinfo {author} {\bibfnamefont {D.~W.}\ \bibnamefont
  {Hobill}},\ }\href@noop {} {\  (\bibinfo {year} {2016})},\ \Eprint
  {http://arxiv.org/abs/1601.06337} {arXiv:1601.06337 [gr-qc]} \BibitemShut
  {NoStop}%
\bibitem [{\citenamefont {Tsui}\ and\ \citenamefont
  {Leung}(2005{\natexlab{a}})}]{PhysRevLett.95.151101}%
  \BibitemOpen
  \bibfield  {author} {\bibinfo {author} {\bibfnamefont {L.~K.}\ \bibnamefont
  {Tsui}}\ and\ \bibinfo {author} {\bibfnamefont {P.~T.}\ \bibnamefont
  {Leung}},\ }\href {\doibase 10.1103/PhysRevLett.95.151101} {\bibfield
  {journal} {\bibinfo  {journal} {Phys. Rev. Lett.}\ }\textbf {\bibinfo
  {volume} {95}},\ \bibinfo {pages} {151101} (\bibinfo {year}
  {2005}{\natexlab{a}})}\BibitemShut {NoStop}%
\bibitem [{\citenamefont {Tsui}\ and\ \citenamefont
  {Leung}(2005{\natexlab{b}})}]{0004-637X-631-1-495}%
  \BibitemOpen
  \bibfield  {author} {\bibinfo {author} {\bibfnamefont {L.~K.}\ \bibnamefont
  {Tsui}}\ and\ \bibinfo {author} {\bibfnamefont {P.~T.}\ \bibnamefont
  {Leung}},\ }\href {http://stacks.iop.org/0004-637X/631/i=1/a=495} {\bibfield
  {journal} {\bibinfo  {journal} {The Astrophysical Journal}\ }\textbf
  {\bibinfo {volume} {631}},\ \bibinfo {pages} {495} (\bibinfo {year}
  {2005}{\natexlab{b}})}\BibitemShut {NoStop}%
\bibitem [{\citenamefont {Tsui}\ \emph {et~al.}(2006)\citenamefont {Tsui},
  \citenamefont {Leung},\ and\ \citenamefont {Wu}}]{Tsui:2006tr}%
  \BibitemOpen
  \bibfield  {author} {\bibinfo {author} {\bibfnamefont {L.~K.}\ \bibnamefont
  {Tsui}}, \bibinfo {author} {\bibfnamefont {P.~T.}\ \bibnamefont {Leung}}, \
  and\ \bibinfo {author} {\bibfnamefont {J.}~\bibnamefont {Wu}},\ }\href
  {\doibase 10.1103/PhysRevD.74.124025} {\bibfield  {journal} {\bibinfo
  {journal} {Phys. Rev.}\ }\textbf {\bibinfo {volume} {D74}},\ \bibinfo {pages}
  {124025} (\bibinfo {year} {2006})},\ \Eprint
  {http://arxiv.org/abs/gr-qc/0610099} {arXiv:gr-qc/0610099 [gr-qc]}
  \BibitemShut {NoStop}%
\bibitem [{\citenamefont {Bowers}\ and\ \citenamefont
  {Liang}(1974)}]{Bowers:1974tgi}%
  \BibitemOpen
  \bibfield  {author} {\bibinfo {author} {\bibfnamefont {R.~L.}\ \bibnamefont
  {Bowers}}\ and\ \bibinfo {author} {\bibfnamefont {E.~P.~T.}\ \bibnamefont
  {Liang}},\ }\href {\doibase 10.1086/152760} {\bibfield  {journal} {\bibinfo
  {journal} {Astrophys. J.}\ }\textbf {\bibinfo {volume} {188}},\ \bibinfo
  {pages} {657} (\bibinfo {year} {1974})}\BibitemShut {NoStop}%
\bibitem [{\citenamefont {Yagi}\ and\ \citenamefont
  {Yunes}(2016)}]{Yagi:2016ejg}%
  \BibitemOpen
  \bibfield  {author} {\bibinfo {author} {\bibfnamefont {K.}~\bibnamefont
  {Yagi}}\ and\ \bibinfo {author} {\bibfnamefont {N.}~\bibnamefont {Yunes}},\
  }\href {\doibase 10.1088/0264-9381/33/9/095005} {\bibfield  {journal}
  {\bibinfo  {journal} {Class. Quant. Grav.}\ }\textbf {\bibinfo {volume}
  {33}},\ \bibinfo {pages} {095005} (\bibinfo {year} {2016})},\ \Eprint
  {http://arxiv.org/abs/1601.02171} {arXiv:1601.02171 [gr-qc]} \BibitemShut
  {NoStop}%
\bibitem [{\citenamefont {Yagi}\ \emph {et~al.}(2016)\citenamefont {Yagi},
  \citenamefont {Stein},\ and\ \citenamefont {Yunes}}]{Yagi:2015oca}%
  \BibitemOpen
  \bibfield  {author} {\bibinfo {author} {\bibfnamefont {K.}~\bibnamefont
  {Yagi}}, \bibinfo {author} {\bibfnamefont {L.~C.}\ \bibnamefont {Stein}}, \
  and\ \bibinfo {author} {\bibfnamefont {N.}~\bibnamefont {Yunes}},\ }\href
  {\doibase 10.1103/PhysRevD.93.024010} {\bibfield  {journal} {\bibinfo
  {journal} {Phys. Rev.}\ }\textbf {\bibinfo {volume} {D93}},\ \bibinfo {pages}
  {024010} (\bibinfo {year} {2016})},\ \Eprint
  {http://arxiv.org/abs/1510.02152} {arXiv:1510.02152 [gr-qc]} \BibitemShut
  {NoStop}%
\bibitem [{\citenamefont {Akmal}\ \emph {et~al.}(1998)\citenamefont {Akmal},
  \citenamefont {Pandharipande},\ and\ \citenamefont {Ravenhall}}]{AP3}%
  \BibitemOpen
  \bibfield  {author} {\bibinfo {author} {\bibfnamefont {A.}~\bibnamefont
  {Akmal}}, \bibinfo {author} {\bibfnamefont {V.~R.}\ \bibnamefont
  {Pandharipande}}, \ and\ \bibinfo {author} {\bibfnamefont {D.~G.}\
  \bibnamefont {Ravenhall}},\ }\href {\doibase 10.1103/PhysRevC.58.1804}
  {\bibfield  {journal} {\bibinfo  {journal} {Phys. Rev. C}\ }\textbf {\bibinfo
  {volume} {58}},\ \bibinfo {pages} {1804} (\bibinfo {year}
  {1998})}\BibitemShut {NoStop}%
\bibitem [{\citenamefont {{Douchin, F.}}\ and\ \citenamefont {{Haensel,
  P.}}(2001)}]{SLy}%
  \BibitemOpen
  \bibfield  {author} {\bibinfo {author} {\bibnamefont {{Douchin, F.}}}\ and\
  \bibinfo {author} {\bibnamefont {{Haensel, P.}}},\ }\href {\doibase
  10.1051/0004-6361:20011402} {\bibfield  {journal} {\bibinfo  {journal}
  {A\&A}\ }\textbf {\bibinfo {volume} {380}},\ \bibinfo {pages} {151} (\bibinfo
  {year} {2001})}\BibitemShut {NoStop}%
\bibitem [{\citenamefont {Wiringa}\ \emph {et~al.}(1988)\citenamefont
  {Wiringa}, \citenamefont {Fiks},\ and\ \citenamefont {Fabrocini}}]{WFF1}%
  \BibitemOpen
  \bibfield  {author} {\bibinfo {author} {\bibfnamefont {R.~B.}\ \bibnamefont
  {Wiringa}}, \bibinfo {author} {\bibfnamefont {V.}~\bibnamefont {Fiks}}, \
  and\ \bibinfo {author} {\bibfnamefont {A.}~\bibnamefont {Fabrocini}},\ }\href
  {\doibase 10.1103/PhysRevC.38.1010} {\bibfield  {journal} {\bibinfo
  {journal} {Phys. Rev. C}\ }\textbf {\bibinfo {volume} {38}},\ \bibinfo
  {pages} {1010} (\bibinfo {year} {1988})}\BibitemShut {NoStop}%
\bibitem [{\citenamefont {Engvik}\ \emph {et~al.}(1996)\citenamefont {Engvik},
  \citenamefont {Bao}, \citenamefont {Hjorth-Jensen}, \citenamefont {Osnes},\
  and\ \citenamefont {Ostgaard}}]{ENG}%
  \BibitemOpen
  \bibfield  {author} {\bibinfo {author} {\bibfnamefont {L.}~\bibnamefont
  {Engvik}}, \bibinfo {author} {\bibfnamefont {G.}~\bibnamefont {Bao}},
  \bibinfo {author} {\bibfnamefont {M.}~\bibnamefont {Hjorth-Jensen}}, \bibinfo
  {author} {\bibfnamefont {E.}~\bibnamefont {Osnes}}, \ and\ \bibinfo {author}
  {\bibfnamefont {E.}~\bibnamefont {Ostgaard}},\ }\href {\doibase
  10.1086/177827} {\bibfield  {journal} {\bibinfo  {journal} {Astrophys. J.}\
  }\textbf {\bibinfo {volume} {469}},\ \bibinfo {pages} {794} (\bibinfo {year}
  {1996})},\ \Eprint {http://arxiv.org/abs/nucl-th/9509016}
  {arXiv:nucl-th/9509016 [nucl-th]} \BibitemShut {NoStop}%
\bibitem [{\citenamefont {Muther}\ \emph {et~al.}(1987)\citenamefont {Muther},
  \citenamefont {Prakash},\ and\ \citenamefont {Ainsworth}}]{MPA1}%
  \BibitemOpen
  \bibfield  {author} {\bibinfo {author} {\bibfnamefont {H.}~\bibnamefont
  {Muther}}, \bibinfo {author} {\bibfnamefont {M.}~\bibnamefont {Prakash}}, \
  and\ \bibinfo {author} {\bibfnamefont {T.}~\bibnamefont {Ainsworth}},\ }\href
  {\doibase https://doi.org/10.1016/0370-2693(87)91611-X} {\bibfield  {journal}
  {\bibinfo  {journal} {Physics Letters B}\ }\textbf {\bibinfo {volume}
  {199}},\ \bibinfo {pages} {469 } (\bibinfo {year} {1987})}\BibitemShut
  {NoStop}%
\bibitem [{\citenamefont {Muller}\ and\ \citenamefont {Serot}(1996)}]{MS1}%
  \BibitemOpen
  \bibfield  {author} {\bibinfo {author} {\bibfnamefont {H.}~\bibnamefont
  {Muller}}\ and\ \bibinfo {author} {\bibfnamefont {B.~D.}\ \bibnamefont
  {Serot}},\ }\href {\doibase https://doi.org/10.1016/0375-9474(96)00187-X}
  {\bibfield  {journal} {\bibinfo  {journal} {Nuclear Physics A}\ }\textbf
  {\bibinfo {volume} {606}},\ \bibinfo {pages} {508 } (\bibinfo {year}
  {1996})}\BibitemShut {NoStop}%
\bibitem [{\citenamefont {Lattimer}\ and\ \citenamefont {Swesty}(1991)}]{LS}%
  \BibitemOpen
  \bibfield  {author} {\bibinfo {author} {\bibfnamefont {J.~M.}\ \bibnamefont
  {Lattimer}}\ and\ \bibinfo {author} {\bibfnamefont {F.~D.}\ \bibnamefont
  {Swesty}},\ }\href {\doibase https://doi.org/10.1016/0375-9474(91)90452-C}
  {\bibfield  {journal} {\bibinfo  {journal} {Nuclear Physics A}\ }\textbf
  {\bibinfo {volume} {535}},\ \bibinfo {pages} {331 } (\bibinfo {year}
  {1991})}\BibitemShut {NoStop}%
\bibitem [{\citenamefont {Shen}\ \emph {et~al.}(1998)\citenamefont {Shen},
  \citenamefont {Toki}, \citenamefont {Oyamatsu},\ and\ \citenamefont
  {Sumiyoshi}}]{SHEN1998435}%
  \BibitemOpen
  \bibfield  {author} {\bibinfo {author} {\bibfnamefont {H.}~\bibnamefont
  {Shen}}, \bibinfo {author} {\bibfnamefont {H.}~\bibnamefont {Toki}}, \bibinfo
  {author} {\bibfnamefont {K.}~\bibnamefont {Oyamatsu}}, \ and\ \bibinfo
  {author} {\bibfnamefont {K.}~\bibnamefont {Sumiyoshi}},\ }\href {\doibase
  https://doi.org/10.1016/S0375-9474(98)00236-X} {\bibfield  {journal}
  {\bibinfo  {journal} {Nuclear Physics A}\ }\textbf {\bibinfo {volume}
  {637}},\ \bibinfo {pages} {435 } (\bibinfo {year} {1998})}\BibitemShut
  {NoStop}%
\bibitem [{\citenamefont {Ozel}\ \emph
  {et~al.}(2016{\natexlab{c}})\citenamefont {Ozel}, \citenamefont {Psaltis},
  \citenamefont {Arzoumanian}, \citenamefont {Morsink},\ and\ \citenamefont
  {Baubock}}]{NICER_radii}%
  \BibitemOpen
  \bibfield  {author} {\bibinfo {author} {\bibfnamefont {F.}~\bibnamefont
  {Ozel}}, \bibinfo {author} {\bibfnamefont {D.}~\bibnamefont {Psaltis}},
  \bibinfo {author} {\bibfnamefont {Z.}~\bibnamefont {Arzoumanian}}, \bibinfo
  {author} {\bibfnamefont {S.}~\bibnamefont {Morsink}}, \ and\ \bibinfo
  {author} {\bibfnamefont {M.}~\bibnamefont {Baubock}},\ }\href {\doibase
  10.3847/0004-637X/832/1/92} {\bibfield  {journal} {\bibinfo  {journal}
  {Astrophys. J.}\ }\textbf {\bibinfo {volume} {832}},\ \bibinfo {pages} {92}
  (\bibinfo {year} {2016}{\natexlab{c}})},\ \Eprint
  {http://arxiv.org/abs/1512.03067} {arXiv:1512.03067 [astro-ph.HE]}
  \BibitemShut {NoStop}%
\bibitem [{\citenamefont {Antoniadis}\ \emph {et~al.}(2013)\citenamefont
  {Antoniadis}, \citenamefont {Freire}, \citenamefont {Wex}, \citenamefont
  {Tauris}, \citenamefont {Lynch} \emph {et~al.}}]{2.01NS}%
  \BibitemOpen
  \bibfield  {author} {\bibinfo {author} {\bibfnamefont {J.}~\bibnamefont
  {Antoniadis}}, \bibinfo {author} {\bibfnamefont {P.~C.}\ \bibnamefont
  {Freire}}, \bibinfo {author} {\bibfnamefont {N.}~\bibnamefont {Wex}},
  \bibinfo {author} {\bibfnamefont {T.~M.}\ \bibnamefont {Tauris}}, \bibinfo
  {author} {\bibfnamefont {R.~S.}\ \bibnamefont {Lynch}},  \emph {et~al.},\
  }\href {\doibase 10.1126/science.1233232} {\bibfield  {journal} {\bibinfo
  {journal} {Science}\ }\textbf {\bibinfo {volume} {340}},\ \bibinfo {pages}
  {6131} (\bibinfo {year} {2013})},\ \Eprint {http://arxiv.org/abs/1304.6875}
  {arXiv:1304.6875 [astro-ph.HE]} \BibitemShut {NoStop}%
\bibitem [{\citenamefont {Yagi}\ and\ \citenamefont
  {Yunes}(2017{\natexlab{a}})}]{EoSclass}%
  \BibitemOpen
  \bibfield  {author} {\bibinfo {author} {\bibfnamefont {K.}~\bibnamefont
  {Yagi}}\ and\ \bibinfo {author} {\bibfnamefont {N.}~\bibnamefont {Yunes}},\
  }\href {http://stacks.iop.org/0264-9381/34/i=1/a=015006} {\bibfield
  {journal} {\bibinfo  {journal} {Classical and Quantum Gravity}\ }\textbf
  {\bibinfo {volume} {34}},\ \bibinfo {pages} {015006} (\bibinfo {year}
  {2017}{\natexlab{a}})}\BibitemShut {NoStop}%
\bibitem [{\citenamefont {Chan}\ \emph {et~al.}(2016)\citenamefont {Chan},
  \citenamefont {Chan},\ and\ \citenamefont {Leung}}]{Chan:2015iou}%
  \BibitemOpen
  \bibfield  {author} {\bibinfo {author} {\bibfnamefont {T.~K.}\ \bibnamefont
  {Chan}}, \bibinfo {author} {\bibfnamefont {A.~P.~O.}\ \bibnamefont {Chan}}, \
  and\ \bibinfo {author} {\bibfnamefont {P.~T.}\ \bibnamefont {Leung}},\ }\href
  {\doibase 10.1103/PhysRevD.93.024033} {\bibfield  {journal} {\bibinfo
  {journal} {Phys. Rev.}\ }\textbf {\bibinfo {volume} {D93}},\ \bibinfo {pages}
  {024033} (\bibinfo {year} {2016})},\ \Eprint
  {http://arxiv.org/abs/1511.08566} {arXiv:1511.08566 [gr-qc]} \BibitemShut
  {NoStop}%
\bibitem [{\citenamefont {Yagi}\ and\ \citenamefont
  {Yunes}(2013)}]{I-Love-Q-Science}%
  \BibitemOpen
  \bibfield  {author} {\bibinfo {author} {\bibfnamefont {K.}~\bibnamefont
  {Yagi}}\ and\ \bibinfo {author} {\bibfnamefont {N.}~\bibnamefont {Yunes}},\
  }\href {\doibase 10.1126/science.1236462} {\bibfield  {journal} {\bibinfo
  {journal} {Science}\ }\textbf {\bibinfo {volume} {341}},\ \bibinfo {pages}
  {365} (\bibinfo {year} {2013})},\ \Eprint {http://arxiv.org/abs/1302.4499}
  {arXiv:1302.4499 [gr-qc]} \BibitemShut {NoStop}%
\bibitem [{\citenamefont {{Yagi}}\ and\ \citenamefont
  {{Yunes}}(2013)}]{I-Love-Q-PRD}%
  \BibitemOpen
  \bibfield  {author} {\bibinfo {author} {\bibfnamefont {K.}~\bibnamefont
  {{Yagi}}}\ and\ \bibinfo {author} {\bibfnamefont {N.}~\bibnamefont
  {{Yunes}}},\ }\href {\doibase 10.1103/PhysRevD.88.023009} {\bibfield
  {journal} {\bibinfo  {journal} {\prd}\ }\textbf {\bibinfo {volume} {88}},\
  \bibinfo {eid} {023009} (\bibinfo {year} {2013})},\ \Eprint
  {http://arxiv.org/abs/1303.1528} {arXiv:1303.1528 [gr-qc]} \BibitemShut
  {NoStop}%
\bibitem [{\citenamefont {Yagi}\ and\ \citenamefont
  {Yunes}(2017{\natexlab{b}})}]{Yagi:2016bkt}%
  \BibitemOpen
  \bibfield  {author} {\bibinfo {author} {\bibfnamefont {K.}~\bibnamefont
  {Yagi}}\ and\ \bibinfo {author} {\bibfnamefont {N.}~\bibnamefont {Yunes}},\
  }\href {\doibase 10.1016/j.physrep.2017.03.002} {\bibfield  {journal}
  {\bibinfo  {journal} {Phys. Rept.}\ }\textbf {\bibinfo {volume} {681}},\
  \bibinfo {pages} {1} (\bibinfo {year} {2017}{\natexlab{b}})},\ \Eprint
  {http://arxiv.org/abs/1608.02582} {arXiv:1608.02582 [gr-qc]} \BibitemShut
  {NoStop}%
\bibitem [{\citenamefont {Doneva}\ and\ \citenamefont
  {Pappas}(2018)}]{Doneva:2017jop}%
  \BibitemOpen
  \bibfield  {author} {\bibinfo {author} {\bibfnamefont {D.~D.}\ \bibnamefont
  {Doneva}}\ and\ \bibinfo {author} {\bibfnamefont {G.}~\bibnamefont
  {Pappas}},\ }\href {\doibase 10.1007/978-3-319-97616-7_13} {\bibfield
  {journal} {\bibinfo  {journal} {Astrophys. Space Sci. Libr.}\ }\textbf
  {\bibinfo {volume} {457}},\ \bibinfo {pages} {737} (\bibinfo {year}
  {2018})},\ \Eprint {http://arxiv.org/abs/1709.08046} {arXiv:1709.08046
  [gr-qc]} \BibitemShut {NoStop}%
\end{thebibliography}%

\end{document}